\documentclass{elsart}

\usepackage{graphicx}

\usepackage{amssymb}
\begin{document}

\begin{frontmatter}


\title{Quantum phase transitions of correlated electrons in two dimensions}
\author{Subir Sachdev}
\ead{subir.sachdev@yale.edu}
\ead[url]{http://pantheon.yale.edu/{\~\,}subir}

\address{Department of Physics, Yale University,\\ P.O. Box 208120, New
Haven, CT 06520-8120, USA.}

\begin{abstract} We review the theories of a few quantum phase transitions
in two-dimensional correlated electron systems and discuss their
application to the cuprate high temperature superconductors. The
coupled-ladder antiferromagnet displays a transition between the
N\'{e}el state and a spin gap paramagnet with a sharp $S=1$
exciton: we develop a careful argument which eventually
establishes that this transition is described by the familiar O(3)
$\varphi^4$ field theory in 2+1 dimensions. Crucial to this
argument is the role played by the quantum Berry phases. We
illustrate this role in a one-dimensional example where our
results can tested against those known from bosonization. On
bipartite lattices in two dimensions, we study the influence of
Berry phases on the quantum transition involving the loss of
N\'{e}el order, and show that they induce bond-centered charge
order ({\em e.g.} spin Peierls order) in the paramagnetic phase.
We extend this theory of magnetic transitions in insulators to
that between an ordinary $d$-wave superconductor and one with
co-existing spin-density-wave order. Finally, we discuss quantum
transitions between superconductors involving changes in the
Cooper pair wavefunction, as in a transition between $d_{x^2-y^2}$
and $d_{x^2-y^2}+id_{xy}$ superconductors. A mean-field theory for
this transition is provided by the usual BCS theory; however, BCS
theory fails near the critical point, and we present the required
field-theoretic extension. Our discussion includes a  perspective
on the phase diagram of the cuprate superconductors, in the
context of recent experiments and the transitions discussed here.
\end{abstract}

\begin{center}
{\small \tt Lectures at the International Summer School on
Fundamental Problems in Statistical Physics X, August-September
2001, Altenberg, Germany}
\end{center}
\end{frontmatter}

\section{Introduction}
\label{intro}

The theory of quantum phase transitions has emerged in the last
decade as a powerful tool for the description of the
unconventional electronic properties of a variety of correlated
systems. This article will provide an elementary overview of some
theoretical models of quantum phase transitions in two spatial
dimensions, along with a brief discussion of their experimental
applications. One reason for our focus on two dimensions is, of
course, our interest in the properties of the cuprate
superconductors, whose interesting electronic excitations reside
within a CuO$_2$ square lattice. However, not coincidentally, two
is also the spatial dimension in which many of the most natural
class of non-trivial quantum critical points reside; three is the
``upper-critical dimension'' of many critical points, at and above
which the physical properties are adequately described by a mean
field theory (possibly extended with some low order fluctuation
corrections).

This article is addressed to newcomers to the study of quantum
phase transitions who desire an introduction to some of the
experimentally relevant theoretical questions under current
discussion. We will assume the reader is well-versed in the
standard theory of classical critical phenomena, and so our focus
will be on novel aspects of quantum phase transitions, including
the role of quantum Berry phases and of excitations with fermionic
statistics. Earlier reviews by this author~\cite{book,science}
provide detailed discussions of background material which we will
occasionally refer the reader to; most of our discussion here goes
beyond these reviews, and focuses on current theoretical and
experimental developments.

We begin in Section~\ref{ladders} by describing the magnetic
quantum transition in the coupled ladder antiferromagnet. High
precision numerical studies of the critical properties of this
transition are now available, and these serve as a useful check on
the theoretical models. We map the critical fluctuations of the
magnetic N\'{e}el order on to the O(3) $\varphi^4$ field theory in
2+1 dimensions, along with additional Berry phase terms which
reflect the commutation relations of the individual quantum spins.
We illustrate the role of the Berry phases by first considering a
one-dimensional antiferromagnet in Section~\ref{xy}, where the
results can be compared with those obtained by bosonization
methods. We return to our discussion of two-dimensional quantum
spin models in Section~\ref{2d}, where we extend the methods
developed in Section~\ref{xy}: we demonstrate that the Berry
phases induce bond-centered charge order in the paramagnetic phase
of antiferomagnets with full square lattice symmetry, while they
can be safely neglected in the coupled ladder antiferromagnet. The
remaining sections extend these theories to doped antiferromagnets
with mobile charge carriers. Section~\ref{magd} considers magnetic
transitions in superconductors, while Section~\ref{did} describes
transitions between BCS superconductors with differing pair
wavefunctions.

\section{Coupled ladder antiferromagnet}
\label{ladders}

We will begin our discussion by describing the quantum phase
transition in a simple two-dimensional model of
antiferromagnetically coupled $S=1/2$ Heisenberg spins. At the
microscopic level, this model does  not describe the spin
fluctuations in the cuprate superconductors. However, a fairly
subtle argument, which we develop in the following sections and
culminate in Section~\ref{magd}, shows that the universal
properties of the simple critical point described in this section
are identical to those of a magnetic ordering transition in the
cuprate superconductors.  The model presented here also provides a
useful description of other transition metal oxides with a spin
gap~\cite{azuma}.

\begin{figure}
\centerline{\includegraphics[width=3in]{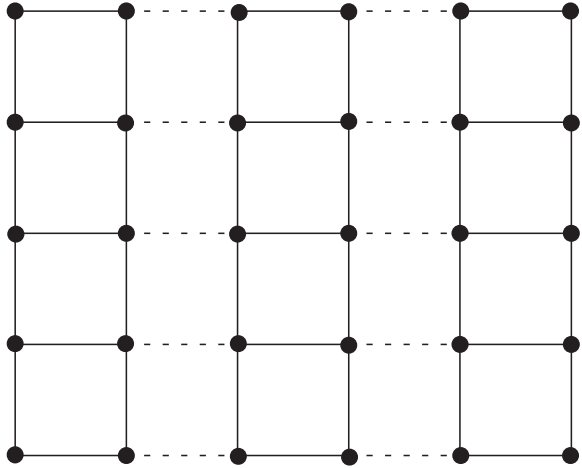}} \caption{ The
coupled ladder antiferromagnet. Spins ($S=1/2$) are placed on the
sites, the $A$ links are shown as full lines, and the $B$ links as
dashed lines.}
\label{fig1}
\end{figure}

We consider the ``ladder'' Hamiltonian
\begin{equation}
H_{\ell} = J \sum_{i,j \in A} {\bf S}_i \cdot {\bf S}_j + \lambda
J \sum_{i,j \in B} {\bf S}_i \cdot {\bf S}_j \label{ham}
\end{equation}
where ${\bf S}_i$ are spin-1/2 operators on the sites of the
coupled-ladder lattice shown in Fig~\ref{fig1}, with the $A$ links
forming decoupled two-leg ladders while the $B$ links couple the
ladders as shown.
The ground state of $H$
depends only on the dimensionless coupling $\lambda$, and we will
describe the low temperature properties as a function of
$\lambda$. We will restrict our attention to $J>0$ and $0 \leq
\lambda \leq 1$.

We will begin with a physical discussion of the phases and
excitations of the coupled ladder antiferromagnet, $H_{\ell}$ in
Section~\ref{sec:lad1}. We will propose a quantum
field-theoretical description of this model in
Section~\ref{sec:lad2}: we will verify that the limiting regimes
of the field theory contain excitations whose quantum numbers are
in accord with the phases discussed in Section~\ref{sec:lad1}, and
will then use the field theory to describe the critical point.

\subsection{Phases and their excitations}
\label{sec:lad1}

Let us first consider the case where $\lambda$ is close to 1.
Exactly at $\lambda=1$, $H$ is identical to the square lattice
Heisenberg antiferromagnet, and this is known to have long-range,
magnetic N\'{e}el order in its ground state {\em i.e.} the
spin-rotation symmetry is broken and the spins have a non-zero,
staggered, expectation value in the ground state with
\begin{equation}
\langle {\bf S}_j \rangle = \eta_j N_0 {\bf n}, \label{neel}
\end{equation}
where ${\bf n}$ is some fixed unit vector in spin space, and
$\eta_j$ is $\pm 1$ on the two sublattices. This long-range order
is expected to be preserved for a finite range of $\lambda$ close
to 1. The low-lying excitations above the ground state consist of
slow spatial deformations in the orientation ${\bf n}$: these are
the familiar spin waves, and they can carry arbitrarily low energy
{\em i.e.} the phase is `gapless'. There are {\em two}
polarizations of spin waves at each wavevector $k = (k_x, k_y)$
(measured from the antiferromagnetic ordering wavevector), and
they have excitation energy $\varepsilon_k =  (c_x^2 k_x^2 + c_y^2
k_y^2)^{1/2}$, with $c_x, c_y$ the spin-wave velocities in the two
spatial directions.

\begin{figure}
\centerline{\includegraphics[width=3.5in]{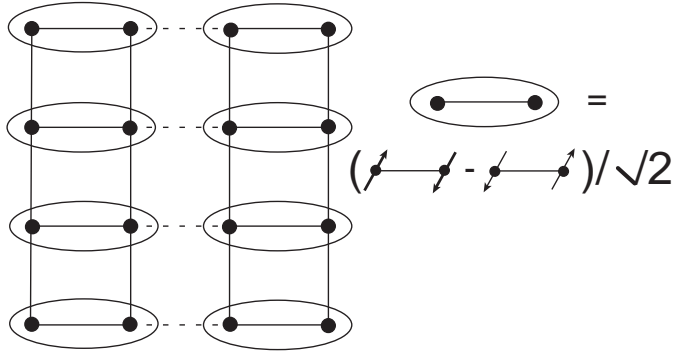}} \caption{
Schematic of the quantum paramagnet ground state for small
$\lambda$. The ovals represent singlet valence bond pairs.
}\label{fig2}
\end{figure}

\begin{figure}
\centerline{\includegraphics[width=5.5in]{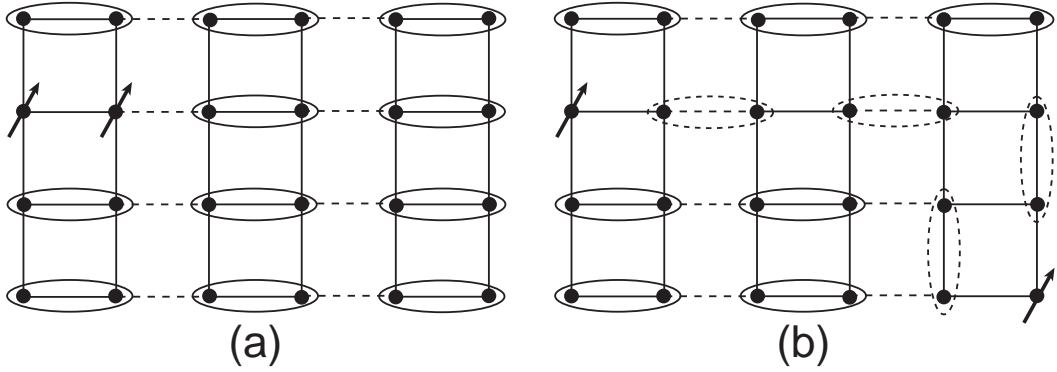}} \caption{(a)
Cartoon picture of the bosonic $S=1$ exciton of the paramagnet.
(b) Fission of the $S=1$ excitation into two $S=1/2$ spinons. The
spinons are connected by a ``string'' of valence bonds (denoted by
dashed ovals) which are not able to resonate with their
environment, and some of which lie on weaker bonds; this string
costs a finite energy per unit length and leads to the confinement
of spinons.} \label{fig3}
\end{figure}

Let us turn now to the vicinity of $\lambda = 0$. Exactly at
$\lambda=0$, $H$ is the Hamiltonian of a set of decoupled spin
ladders. Such spin ladders are known to have a paramagnetic ground
state, with spin rotation symmetry preserved, and an energy gap to
all excitations~\cite{dagotto}. A caricature of the ground state
is sketched in Fig~\ref{fig2}: spins on opposite rungs of the
ladder pair in valence bond singlets in a manner which preserves
all lattice symmetries.
Excitations are now formed by breaking a valence bond, which leads
to a {\em three}-fold degenerate state with total spin $S=1$, as
shown in Fig~\ref{fig3}; this broken bond can hop from
site-to-site, leading to a triplet quasiparticle excitation.
Note that this quasiparticle is {\em not\/} a spin-wave (or
equivalently, a `magnon') but is more properly referred to as a
spin 1 {\em exciton}. For $\lambda$ small, but not exactly 0, we
expect that the ground state will remain a gapped paramagnet, and
the spin exciton will now move in two dimensions. We parameterize
its energy at small wavevectors by
\begin{equation}
\varepsilon_k = \Delta + \frac{c_x^2 k_x^2 + c_y^2 k_y^2}{2
\Delta}, \label{epart}
\end{equation}
where $\Delta$ is the spin gap, and $c_x$, $c_y$ are velocities.
Fig~\ref{fig3} also presents a simple argument which shows that
the $S=1$ exciton cannot fission into two $S=1/2$ `spinons'.

The very distinct symmetry signatures of the ground states and
excitations between $\lambda \approx 1$ and $\lambda \approx 0$
make it clear that the two limits cannot be continuously
connected. It is known~\cite{katoh,twor} that there is an
intermediate second-order phase transition at $\lambda = \lambda_c
\approx 0.3$. Both the spin gap $\Delta$ and the N\'{e}el order
parameter $N_0$ vanish continuously as $\lambda_c$ is approached
from either side.

\subsection{Quantum field theory}
\label{sec:lad2}

A convenient field-theoretic description of the two phases
introduced in Section~\ref{sec:lad1}, and of the critical point
which separates them, is obtained by the coherent state path
integral. A detailed derivation of this approach has been provided
by the author in Chapter 13 of Ref.~\cite{book} which the reader
should now consult; we will only highlight a few important points
here.

The coherent state path integral is best introduced by considering
first a single spin ${\bf S}$ with angular momentum $S$ and
Hamiltonian
\begin{equation}
H_0 = - {\bf h} \cdot {\bf S} \label{h0}
\end{equation}
where ${\bf h}$ is a static magnetic field. The $(2S+1)$
eigenvalues of $H_0$ are, of course, $-m |{\bf h}|$, where $m =
-S,-S+1, \ldots, S-1,S$, and so we also know its partition
function $Z_0$ at a temperature $T$. The coherent state path
integral represents $Z_0$ in terms of an integral over all
possible closed curves, ${\bf N}(\tau)$, on the surface of a unit
sphere, each curve representing a possible history of the
precessing spin in imaginary time, $\tau$:
\begin{equation}
Z_0 = \int \mathcal{D} {\bf N}(\tau) \delta( {\bf N}^2 (\tau) - 1)
\exp \left( - i S \int \mathcal{A}_\tau ({\bf N}(\tau)) d \tau +
\int d \tau S {\bf h} \cdot {\bf N}(\tau) \right). \label{z0}
\end{equation}
The first term is the so-called `Berry phase', and
$\mathcal{A}_\tau ({\bf N}(\tau)) d \tau$ is defined to be the
oriented area of the spherical triangle defined by ${\bf N}
(\tau)$, ${\bf N} (\tau + d \tau)$, and an arbitrary reference
point ${\bf N}_0$ (which is usually chosen to be the north pole).
We will see that $\mathcal{A}_{\tau}$ behaves in many respects
like the time-component of a gauge field, and this accounts for
the suggestive notation. All physical results should be
independent of the choice of ${\bf N}_0$, and it is easy to see
that changes in ${\bf N}_0$ amount to gauge transformations of
$\mathcal{A}_{\tau}$. So if we change ${\bf N}_0$ to ${\bf
N}_0^{\prime}$, then the resulting $\mathcal{A}_{\tau}^{\prime}$
is related to $\mathcal{A}_{\tau}$ by
\begin{equation}
\mathcal{A}_{\tau}^{\prime} = \mathcal{A}_{\tau} - \partial_{\tau}
\phi (\tau) \label{gauge}
\end{equation}
where $\phi(\tau)$ measures the oriented area of the spherical
triangle defined by ${\bf N} (\tau)$, ${\bf N}_0$, and ${\bf
N}_0^{\prime}$.

We now apply the above path integral independently to every site
of the coupled ladder antiferromagnet: we obtain a path integral
over fields ${\bf N}_j (\tau)$, where $j$ is a site index as in
(\ref{ham}). As discussed in Chapter 13 of Ref.~\cite{book}, it is
convenient to now take a spatial continuum limit, to allow focus
on the important low energy excitations. This proceeds by the
parameterization
\begin{equation}
{\bf N}_j (\tau) = \eta_j {\bf n} (r_j, \tau) \sqrt{1 - (a^2/S)^2
{\bf L}^2 (r_j ,\tau)} + \frac{a^2}{S} {\bf L} (r_j,\tau),
\label{cont}
\end{equation}
where $\eta_j = \pm 1$ on the two subattices, $a$ is the lattice
spacing, $r \equiv (x,y)$, ${\bf n} (r,\tau)$ is a unit length
continuum field representing the staggered N\'{e}el order
parameter, and ${\bf L} (r, \tau)$ is the continuum field for the
conserved, uniform, magnetization density (the normalization
chosen in (\ref{cont}) is that required to make $\int d^2 r {\bf
L} (r, \tau)$ exactly equal to the total spin). Next, a gradient
expansion in performed on the effective action for $H_{\ell}$, and
after integrating out the ${\bf L}$ fluctuations we obtain the
following path integral for the partition function of the coupled
ladder antiferromagnet:
\begin{eqnarray}
Z_{\ell} &=& \int \mathcal{D} {\bf n} (r, \tau) \delta ( {\bf n}^2
(r, \tau) - 1) \exp \Biggl[ - i S \sum_j \eta_j \int d \tau
\mathcal{A}_{\tau} ({\bf n}(r_j, \tau)) \nonumber \\
&~&~~~~~~~~~~- \frac{1}{2 g \sqrt{c_x c_y}} \int d^2 r d\tau
\left( (\partial_{\tau} {\bf n})^2 + c_x^2 (\partial_x {\bf n})^2
+ c_y^2 (\partial_y {\bf n})^2 \right) \Biggr], \label{zl}
\end{eqnarray}
where the velocities $c_x = JSa\sqrt{(1+\lambda)(3+\lambda)} $,
$c_y = JSa \sqrt{2(3+\lambda)} $, and the coupling constant $g =
(2a/S) [(3+\lambda)^2/(2 + 2 \lambda)]^{1/4}$. We will now view
$g$ as the tuning parameter, and values of $g$ on either side of a
critical value $g_c$  place the system in one of the two phases
discussed in Section~\ref{sec:lad1}; notice that $g$ is a
decreasing function of $\lambda$ for $0 < \lambda < 1$, and so the
``strong'' coupling phase $g>g_c$ corresponds to the spin gap
phase with $\lambda < \lambda_c$, while $g<g_c$ is the $\lambda >
\lambda_c$ N\'{e}el phase.

The first term in (\ref{zl}) is a residual Berry phase, and this
remains expressed on the underlying lattice: there is no natural
way to take its continuum limit.  Notice that this term induces
complex weights in the partition function and so it has no analog
in the theory of classical phase transitions. This Berry phase
plays a central role in determining the phases and critical points
of many quantum spin systems, and describing its consequences is
one of the central purposes of this article. However, our
discussions in the following sections will show that at least for
the case of the coupled ladder antiferromagnet the Berry phase in
(\ref{zl}) can be safely neglected, and so we will drop it in the
remainder of this subsection.

Upon omission of the Berry phase in (\ref{zl}), $Z_{\ell}$ becomes
a standard path integral, familiar in well established theories of
classical phase transitions in $3=2+1$ dimensions. For instance,
after interpreting $\tau$ as a third dimension ({\em i.e.\/} $\tau
\rightarrow z$), and ${\bf n}$ as the local magnetization in a
three-dimensional ferromagnet ({\em e.g.} iron), then $Z_{\ell}$
is simply its {\em classical} partition function at a finite
`temperature' $T_{cl}$, with the coupling constant $g \sim
T_{cl}$. From the theory of such ferromagnets it is known that a
convenient description of the physical properties of $Z_{\ell}$
emerges not in a formulation in terms of the fixed length field
${\bf n}$ (the `hard spin' representation), but in a formulation
using a `soft spin' field $\varphi_{\alpha} (r, \tau)$ ($\alpha =
x,y,z$) with no length constraint. We connect the two formulations
by a coarse-graining transformation, so that $\varphi_{\alpha}$
represents the average of ${\bf n}$ over some small
coarse-graining volume of spacetime. The new form of $Z_{\ell}$ is
now
\begin{eqnarray}
Z_{\ell} = \int \mathcal{D} \varphi_{\alpha} (r, \tau) \exp
\Biggl[ - \int d^2 r d \tau && \left\{ \frac{1}{2} \left(
(\partial_{\tau} \varphi_{\alpha} )^2 + c_x^2 (\partial_x
\varphi_{\alpha} )^2 + c_y^2 (\partial_y \varphi_{\alpha} )^2
\right.\right. \nonumber \\
&&~~~~~~~~\left.\left. +s \varphi_{\alpha}^2 \right) +
\frac{u}{24} \left( \varphi_{\alpha}^2 \right)^2 \right\} \Biggr],
\label{zl1}
\end{eqnarray}
where keep in mind that the Berry phase in (\ref{zl}) has been
dropped, $s$ is now the new tuning parameter (replacing $g$ in
(\ref{zl})), and the quartic interaction $u$ has replaced the
fixed length constraint in (\ref{zl}). The two phases of
Section~\ref{sec:lad1} are now separated by a critical value
$s=s_c$, with the $\lambda < \lambda_c$ spin gap phase
corresponding to $s > s_c$, while the $\lambda> \lambda_c$
N\'{e}el phase appears for $s < s_c$.

The well-known properties of the model (\ref{zl1}), when
translated into the quantum framework of interest here, are easily
seen to reproduce the physical properties of the phases discussed
in Section~\ref{sec:lad1}.

For $s<s_c$, $Z_{\ell}$ in (\ref{zl1}) has a ``ferromagnetic''
phase associated with the breaking of spin rotation invariance,
and the development of the expectation value $\langle
\varphi_{\alpha} \rangle = \delta_{\alpha z} N_0$; to lowest order
in $u$, the spontaneous magnetization is $N_0 = \sqrt{-6s/u}$ and
is determined by the minimum of the ``Mexican hat'' potential for
the $\varphi_{\alpha}$ field. After using (\ref{cont}) this
``ferromagnetic'' order in the classical model is seen to
correspond to the antiferromagnetic N\'{e}el order (\ref{neel}) in
the quantum model. A standard Gaussian fluctuation analysis about
this ordered states shows that susceptiblity transverse to the
spontaneous moment is given by
\begin{equation}
\chi_{\perp} (k, \omega_n) = \left\langle \left| \varphi_x (k,
\omega_n) \right|^2 \right\rangle = \left\langle \left| \varphi_y
(k, \omega_n) \right|^2 \right\rangle = \frac{1}{\omega_n^2 +
c_x^2 k_x^2 + c_y k_y^2} \label{e1}
\end{equation}
where $\omega_n$ is a frequency associated with imaginary time.
After analytically continuing to real frequencies, $\omega$, this
yields the quantum response function to an applied staggered field
transverse to the N\'{e}el order:
\begin{equation}
\chi_{\perp} (k, \omega) = \frac{1}{c_x^2 k_x^2 + c_y k_y^2 -
\omega^2}. \label{e2}
\end{equation}
This response function has a pole, representing the two spin-wave
excitations, at energy $\left( c_x^2 k_x^2 + c_y^2 k_y^2
\right)^{1/2}$. This is in complete accord with our discussion of
the properties of the N\'{e}el phase in Section~\ref{sec:lad1}.

For $s>s_c$, the ``disordered'' phase of $Z_{\ell}$ in (\ref{zl1})
corresponds to the quantum paramagnet with a spin gap. Here, a
computation of the susceptibility to lowest order in $u$ yields
\begin{equation}
\chi (k, \omega) = \frac{\mathcal{Z}}{ \Delta^2 + c_x^2 k_x^2 +
c_y k_y^2 - \omega^2}, \label{e3}
\end{equation}
where we have already analytically continued to real frequencies,
$\Delta=\sqrt{s}$, and (for now) the residue $\mathcal{Z}=1$. This
response function is independent of the direction of the applied
staggered field, and so its pole represents the $S=1$ exciton at
energy $\left(\Delta^2+ c_x^2 k_x^2 + c_y^2 k_y^2 \right)^{1/2}$.
This is again in accord with the discussion of the spin gap phase
in Section~\ref{sec:lad1}: the exciton dispersion (\ref{epart}) is
obtained from the present result in a small momentum expansion.
The pole structure in (\ref{e3}) actually holds to all orders in
an expansion in $u$~\cite{book}: at quantum temperature $T=0$,
such corrections renormalize the values of the spin gap $\Delta$
and the quasiparticle residue $\mathcal{Z}$, and induce additional
spectral weight associated with multiple excitonic continua at
energies larger than $3 \Delta$, but they do not broaden the
quasiparticle pole. Physically, this happens because the exciton
is the lowest energy excitation with spin $S=1$, and so
conservation of total spin prevents implies that this excitation
has an infinite lifetime. Only at $T>0$ does a finite lifetime
appear, associated with scattering off the thermally excited
density of pre-exiting excitons.

The present field theoretical framework can also be extended to
obtain a theory for the critical point at $s=s_c$
($\lambda=\lambda_c$), as has been discussed in some detail
in~\cite{book}. In its application to the classical ferromagnet,
the critical properties of (\ref{zl1}) have been understood in
some detail in studies of the $\varphi^4$ field theory, and
accurate values of the critical exponents have been obtained. In a
recent quantum Monte Carlo study of the coupled ladder
antiferromagnet $H_{\ell}$, Matsumoto {\em et al.}~\cite{matsu}
determined the critical exponents at $\lambda=\lambda_c$ to
impressive accuracy, and all values are in accord with the
exponents of the $\varphi^4$ field theory. This agreement is
independent numerical support for the omission of the Berry phases
in proceeding from (\ref{zl}) to (\ref{zl1}); separate theoretical
arguments for the same approximation will emerge in the following
sections. As we approach the critical point with $s \searrow s_c$,
the pole structure in (\ref{e3}) continues to hold at low
frequencies, but the spin gap vanishes as $\Delta \sim (s-s_c)^{z
\nu}$ and the residue vanishes as $\mathcal{Z} \sim (s-s_c)^{\eta
\nu}$, where $z=1$, $\nu$, and $\eta$ are standard critical
exponents. Right at $s=s_c$ there are no quasiparticle excitions,
and the dynamic susceptiblity instead displays a branch-cut
representing the continuum of critical excitations:
\begin{equation}
\chi (k, \omega ) \sim \frac{1}{\left( c_x^2 k_x^2 + c_y^2 k_y^2 -
\omega^2 \right)^{1-\eta/2}}. \label{e4}
\end{equation}
For more information on the physics of this critical continuum,
refer to~\cite{book}.

\section{Quantum XY chain}
\label{xy}

This section will divert from the main subject of this paper by
considering some simple examples of quantum phase transitions in
one dimension. Our purpose here is to illustrate the consequences
of the Berry phases in (\ref{zl}) by considering a simplified
model for which they can be evaluated completely; we will also be
able to test our method by comparing the results with those
obtained by a bosonization analysis of the same one-dimensional
model. We will then proceed to apply closely related methods in
two dimensions in Section~\ref{2d}.

We will evaluate the generalization of (\ref{zl}) in one dimension
for case in which the Hamiltonian has on a global U(1) spin
symmetry, and the anisotropies are such that the spins prefer to
lie within the $x$-$y$ plane in spin space. In this case we can
parameterize ${\bf n}$ by a single angle $\theta$:
\begin{equation}
{\bf n} = (\cos \theta, \sin \theta, 0); \label{e5}
\end{equation}
with these simplifications, we will find that powerful duality
methods enable us to obtain considerable insight into the physics
a partition function with Berry phases and complex weights. With
the parameterization (\ref{e5}) and the restriction to one spatial
dimension, we expect the resulting $Z_{\ell}$ to provide a
description of a large class of spin $S$ quantum antiferromagnets
in one dimension; to be specfic, we expect it will model the
Hamiltonian
\begin{eqnarray}
H_{XY} &=& J_1 \sum_j \left( S_{xj} S_{x,j+1} + S_{yj} S_{y,j+1}+
\zeta S_{zj} S_{z,j+1} \right) \nonumber \\ &~&~~~~~~~~~~~~~~~+
J_2 \sum_j \left( S_{xj} S_{x,j+2} + S_{yj} S_{y,j+2}+ \zeta
S_{zj} S_{z,j+2}\right), \label{e6}
\end{eqnarray}
where $J_{1,2} > 0$ are first/second neighbor antiferromagnetic
exchange constants ($J_2$ is not too large), and the anisotropy
$|\zeta| < 1$ induces a preference for the spins to reside in the
$x$-$y$ plane. We can reasonably expect that the phase diagram of
$H_{XY}$ as a function of $\zeta$ and $J_2/J_1$ is similar to that
of $Z_{\ell}$ as a function of $g$. The former is already known by
an earlier bosonization analysis~\cite{xxz1d} and will provide an
instructive comparison with the latter.

\begin{figure}
\centerline{\includegraphics[width=3in]{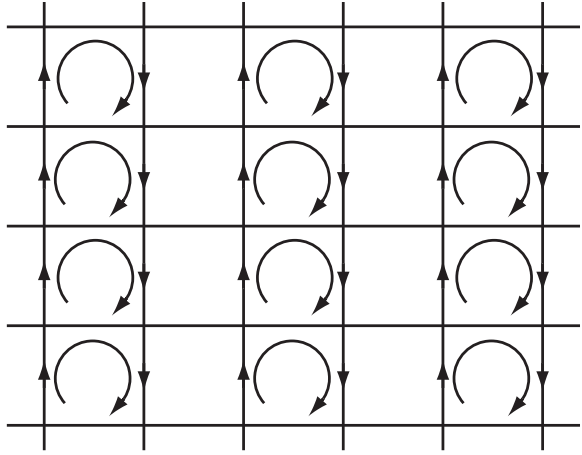}}
\caption{Diagrammatic proof of the identity (\protect\ref{e7})
which relates the Berry phase to the sum over the
$\mathcal{A}_{\mu}$ flux in every plaquette in the odd columns.}
\label{fig4}
\end{figure}

Our computation begins by a transformation of the Berry phase in
(\ref{zl}) to an alternative form in one dimension. We discretize
spacetime into a square lattice of sites, $j$, ($j$ is now a
spacetime index, unlike the spatial index in (\ref{e6}), and its
interpretation should be clear from the context). For this
spacetime lattice we define $\mathcal{A}_{j \mu}$, with
$\mu=x,\tau$, to be the oriented area of the spherical triangle
formed by ${\bf n}_j$, ${\bf n}_{j+\hat{\mu}}$, and a fixed
reference ${\bf N}_0$. Then, as illustrated in Fig~\ref{fig4}, we
can use the identity
\begin{equation}
\sum_j \eta_j \mathcal{A}_{j \tau} = \sum_j \ell_{\bar{\jmath}}
\epsilon_{\mu\nu} \Delta_{\mu} \mathcal{A}_{j\nu} \label{e7}
\end{equation}
to relate the Berry phase to the $\mathcal{A}_{\mu}$ flux piercing
the plaquettes of the spacetime lattice;
here $\ell_{\bar{\jmath}}$ is a fixed integer field on the sites,
$\bar{\jmath}$, of the dual lattice, and a convenient choice is
$\ell_{\bar{\jmath}} = (1 - (-1)^{\bar{\jmath}_x} )/2$ {\em i.e.}
$\ell_{\bar{\jmath}}$ is unity on every odd column, and zero on
every even column. The symbol $\Delta_{\mu}$ represents a discrete
lattice derivative, with $\Delta_{\mu} f_j \equiv f_{j +
\hat{\mu}} - f_j$.

Now we use the in-plane form of ${\bf n}$ in (\ref{e5}) to our
advantage. The flux $\epsilon_{\mu\nu} \Delta_{\mu} \mathcal{A}_{j
\nu}$ through any plaquette is clearly equal to the surface area
of the quadrilateral on the unit sphere formed by the ${\bf n}_j$
on the four vertices of the plaquette. For planar spins, such a
quadrilateral lies entirely on the equator, and its area is
generically zero. The only exception occurs when a vortex in
$\theta$ pierces the plaquette: in this case, the quadrilateral is
the entire equator and the flux is the area of a hemisphere $= \pm
2 \pi$. So we have established that the Berry phases in (\ref{zl})
generate the factor
\begin{equation}
\exp \left( - i 2 \pi S \sum_{\bar{\jmath}} \ell_{\bar{\jmath}}
\times (\mbox{vortex number of plaquette $\bar{\jmath}$}) \right)
\label{e8}
\end{equation}
for easy-plane spin systems in one dimension. As the vortex number
is always an integer, we see that the factor in (\ref{e8}) is
always unity for integer $S$, and the Berry phases can be ignored
for these values of $S$.

The remaining terms in the action of (\ref{zl}) can also be
simplified for the present situation. On the discrete spacetime
lattice, and with (\ref{e5}), these terms become $-(1/g) \sum_{j}
\cos (\Delta_{\mu} \theta_j )$, which is the usual action of a
``classical'' XY model in $2=1+1$ dimensions. As is standard,
duality manipulations on this XY model are more conveniently
performed in a periodic Gaussian, `Villain', representation of the
weights of the model. In this manner, we finally obtain the
generalization of the partition function (\ref{zl}) to U(1)
symmetric, quantum spin $S$ chains:
\begin{equation}
Z_{XY} = \sum_{\{m_{j \mu}\}} \int \prod_j d \theta_j \exp \left(
- \frac{1}{2g} \sum_{j\mu} ( \Delta_{\mu} \theta_j - 2 \pi m_{j
\mu})^2 -  i 2 \pi S \sum_{\Box} \epsilon_{\mu\nu}
\ell_{\bar{\jmath}} \Delta_{\mu} m_{j \nu} \right), \label{e9}
\end{equation}
where $m_{j \mu}$ are integers on the links of the direct lattice
which, when summed, produce the periodic Gaussian weights for
$\Delta_{\mu} \theta_j$, the $\Box$ represents a sum over
plaquettes, and $\epsilon_{\mu\nu} \Delta_{\mu} m_{j \nu}$ are the
vortex numbers of the plaquettes. The expression (\ref{e9}) is
amenable to an exact duality transformation which produces a
representation in which all weights are positive: we apply the
Poisson summation identity to the sum over $m_{j \mu}$ on each
link, and replace it by a sum over integers $J_{j \mu}$:
\begin{equation}
Z_{XY} = \sum_{\{J_{j \mu}\}} \int \prod_j d \theta_j \exp \left(
- \frac{g}{2} \sum_{j\mu} ( J_{j \mu} - S \epsilon_{\mu\nu}
\Delta_{\nu} \ell_{\bar{\jmath}})^2 - i \sum_{j\mu} J_{j \mu}
\Delta_{\mu} \theta_j \right). \label{e9a}
\end{equation}
Here, and henceforth, we drop overall normalization constants in
front of partition functions. Now the integral over the $\theta_j$
can be performed independently on each site, and only imposes the
constraint $\Delta_{\mu} J_{j \mu} = 0$. We solve this constraint
by writing $J_{j \mu} = \epsilon_{\mu\nu} \Delta_{\nu}
p_{\bar{\jmath}}$, with the $p_{\bar{\jmath}}$ integers on the
sites of the dual lattice. So we obtain the exact mapping of
(\ref{e9}) to~\cite{kwon}
\begin{equation}
Z_{XY} = \sum_{\{H_{\bar{\jmath}}\}} \exp \left( -\frac{g}{2}
\sum_{\bar{\jmath}} \left(\Delta_{\mu} H_{\bar{\jmath}} \right)^2
\right), \label{e10}
\end{equation}
where the ``heights'' $H_{\bar{\jmath}}$ on the sites of the dual
lattice  are restricted to the values
\begin{equation}
H_{\bar{\jmath}} \equiv p_{\bar{\jmath}} - S \ell_{\bar{\jmath}}
\label{e10a}
\end{equation}
The expression (\ref{e10}) is a canonical height (or
`solid-on-solid') model in two dimensions. The only influence of
the quantum Berry phase are the restrictions on the allowed values
of the heights of the ``interface'': for integer $S$ the heights
take all integer values just as in the standard solid-on-solid
model, while for half-integer $S$, the heights are restricted to
half-odd-integer (integer) values on the odd (even) columns of the
square lattice. As we have made a number of approximations in the
deriving $Z_{XY}$ from the path integral of spin systems, we shall
be interested in a generalized class of height models with
additional short-range non-local couplings, apart from the simple
terms appearing in (\ref{e10}). With these additional couplings,
standard methods from the theory of interface models can be used
to show~\cite{kwon} that the generalized height models describe
phases not only in the easy-plane, small $|\zeta|$, limit of
(\ref{e6}), but also those in the easy-axis, large $|\zeta|$ limit
when the spins prefer to lie in the $\pm z$ directions. In all of
this analysis it is essential that we not relax the restrictions
on the allowed values of the heights, as these are fundamental
consequences of the quantized value of the spin $S$. We describe
the main results for two sets of $S$ values in turn.

\subsection{$S$ integer}
\label{int}

The standard model with integer heights on every site has two
phases. For small $g$, the interface is rough; this is dual to the
classical or quantum XY phase with quasi-long-range spin order in
the $x$-$y$ plane, with power-law decay of spin correlations. For
the quantum spin chain, we may refer to this phase as a
Tomonaga-Luttinger (TL) liquid, as it shares many of the
characteristics of other TL liquids.

For large $g$, the interface is smooth; this is dual to the
``disordered'' phase of the classical XY model. For the quantum
spin chain, this large $g$ phase has exponential decay of spin
correlations in imaginary time, and therefore a spin gap. Its
characteristics are rather similar to the spin gap phase discussed
in Section~\ref{ladders}: the primary excitation is a gapped $S=1$
exciton with an infinite lifetime at low energies. As the spin
anisotropy is reduced by increasing $\zeta$ in (\ref{e6}), this
spin gap phase connects continuously to the well-known Haldane gap
state of integer spin chains with SU(2) symmetry.

The transition between the small and large $g$ phases is
second-order, and is described by the standard Kosterlitz-Thouless
theory of the classical XY model in two dimensions.

\subsection{$S$ half-odd-integer}
\label{hint}

The staggering of the heights between integer and half-odd-integer
values introduces crucial differences in the phase diagram, and
several new phases are allowed in the generalized phase diagram of
such height models. However, it is believed~\cite{kwon} that the
specific height model (\ref{e10}), with only a local gradient
squared coupling, has only one rough phase for all values of $g$,
and this corresponds, as above, to a TL phase of the quantum spin
chain. The other phases appear upon allowing additional
short-range couplings in (\ref{e10}), and these can reasonably be
expected to arise from terms in the original Hamiltonian omitted
in the duality mappings; as we noted above, in this generalization
it is essential that we maintain the restrictions on the allowed
height values. The new phases are associated with smooth
interfaces, and as in Section~\ref{int} such phases have a spin
gap. However, the staggering of the heights induces crucial
differences in the structure of such spin gap states, as we
describe below.

\begin{figure}
\centerline{\includegraphics[width=5in]{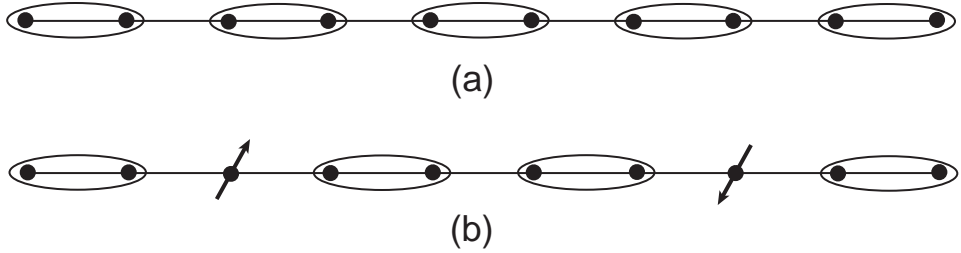}} \caption{(a)
Schematic of the state with bond-centered charge (BC) or
spin-Peierls order for the $S=1/2$ quantum spin chain. The
ellipses represent valence bonds as in Fig~\protect\ref{fig2}, and
these have spontaneously chosen one of two possible dimerizations
of the chain. (b) $S=1/2$ spin excitations of the BC state. Unlike
the case in Fig~\protect\ref{fig3}b, these spinons are free to
move far apart from each other: this is because we are in one
dimension {\em and} the dimerization is spontaneous. Such spinons
are confined in two dimensions whether the BC order is spontaneous
or explicitly present in the Hamiltonian.} \label{fig5}
\end{figure}

A crucial characteristic of a smooth interface is that there is a
well-defined value of the average height $\langle H_{\bar{\jmath}}
\rangle$ in every pure state. Furthermore, for the staggered
height models under consideration here, every value of the average
$\langle H_{\bar{\jmath}} \rangle$ {\em breaks a lattice
translational symmetry of the quantum spin chain}. For instance if
$\langle H_{\bar{\jmath}} \rangle = 0$ (modulo integers) then the
even columns of the square lattice (which have integer allowed
heights) have been picked out, and we may expect the average
energy, $\langle S_{jx}S_{j+1,x} + S_{jy} S_{j+1,y} \rangle$ on
the links associated with the even columns to be smaller (say)
than those on the odd columns. Conversely, an interface with
$\langle H_{\bar{\jmath}} \rangle = 1/2$ (modulo integers)
reverses the role of the odd and even columns. The choice between
these two sets of average heights corresponds to a spontaneously
broken $Z_2$ symmetry in the ground state, which now has
`spin-Peierls' or `bond-centered charge' (BC) order. The nature of
these ordered states, and of their excitations, is illustrated in
Fig~\ref{fig5}: they have a spin gap, but the lowest-lying
excitations are fractionalized $S=1/2$ spinons which reside on the
domain wall between the two BC states.

Other possible values of $\langle H_{\bar{\jmath}} \rangle$ in the
smooth interface phase lead to other discrete broken symmetries in
the quantum spin chain. As we describe in Ref.~\cite{kwon}, the
values $\langle H_{\bar{\jmath}} \rangle = 1/4,3/4$ (modulo
integers) correspond to states with Ising spin order {\em i.e.\/}
$\langle S_{jz} \rangle = \pm \eta_j N_0$. All other values of
$\langle H_{\bar{\jmath}} \rangle$ correspond to four-fold
degenerate states with co-existing Ising and BC order.

\begin{figure}
\centerline{\includegraphics[width=3in]{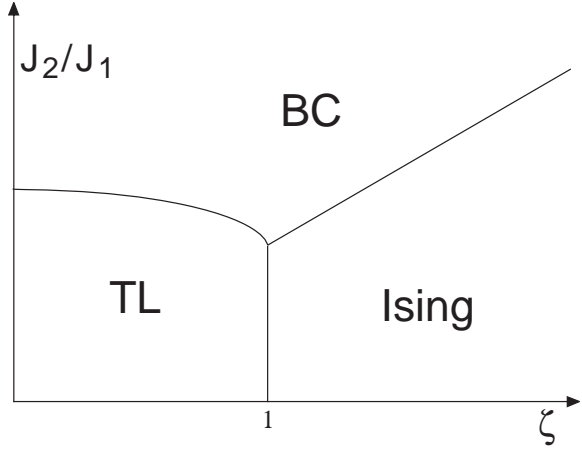}} \caption{Phase
diagram of the $S=1/2$ quantum spin chain $H_{XY}$ as obtained by
abelian bosonization \protect\cite{xxz1d}. Identical results are
obtained from a generalized square lattice height model in which
the heights take half-odd-integer (integer) values on the even
(odd) columns.} \label{fig6}
\end{figure}

We sketch the phase diagram of the model (\ref{e10}) for $S$
half-odd-integer in Fig~\ref{fig6}, as obtained by a bosonization
analysis~\cite{xxz1d}.
The TL, BC, and Ising phases are all separated by second-order
transition lines, and the three lines meet at a single
multi-critical point: all the critical properties can be described
by an extension of the Kosterlitz-Thouless renormalization group
flows. Analysis of the generalized height models discussed here
leads to a phase diagram with an identical topology and critical
properties \cite{kwon}.

\section{Berry phases in two dimensions}
\label{2d}

We are now prepared to describe the evaluation of the 2+1
dimensional path integral with Berry phases in (\ref{zl}). As in
the 1+1 dimensional case in Section~\ref{xy} the evaluation is
performed on a discrete lattice in spacetime: the degrees of
freedom are three component unit vectors, ${\bf n}_j$, on the
sites, $j$ of a cubic lattice. On such a lattice we can rewrite
(\ref{zl}) as
\begin{equation}
Z = \int \prod_j d {\bf n}_j \delta({\bf n}_j^2 - 1) \exp \left( -
\frac{1}{2g} \sum_{j,\mu} {\bf n}_j \cdot {\bf n}_{j + \hat{\mu}}
- i S \sum_j \eta_j \mathcal{A}_{j\tau} \right), \label{f1}
\end{equation}
where the sum over $\mu$ extends over the three directions
$x$,$y$,$\tau$. In the application to the coupled latter
antiferromagnet, the spatial anisotropy and doubled unit cell of
$H_{\ell}$ led to an anisotropy in the spatial derivative terms in
(\ref{zl}). In the discrete form (\ref{f1}) this reduced symmetry
requires a corresponding variation in the couplings $g$ associated
with the couplings between the various sites on the cubic
spacetime lattice: we have dropped this variation in (\ref{f1})
and so, strictly speaking, this partition function applies only to
quantum antiferromagnets which have the full symmetry of the
square lattice and have a single spin $S$ on each site. An example
of such an antiferromagnet is the so-called $J_1$-$J_2$ model
\begin{equation}
H = J_1 \sum_{\langle ij \rangle} {\bf S}_i \cdot {\bf S}_j + J_2
\sum_{{\rm nnn}~ij} {\bf S}_i \cdot {\bf S}_j \label{f2}
\end{equation}
with exchange interaction $J_1>0$ between square lattice nearest
neighbors $\langle ij \rangle$, and $J_2 >0$ between
next-nearest-neighbors (nnn). As in Section~\ref{xy}, we exclude
large values of $J_2$ and expect that the phases of $H$ as a
function of $J_2 /J_1$ will be described by those of $Z$ in
(\ref{f1}) as a function of $g$. Later we will also discuss the
consequences of doubling the unit cell and introducing spatial
anisotropy in (\ref{f1}), which is required to obtain a
description of the coupled ladder model in (\ref{ham}).

For small $g$, we expect $Z$ to be in the ordered N\'{e}el state,
with broken spin rotation symmetry. It is not difficult to account
for the Berry phases in a spin-wave fluctuation analysis about
this ordered state, and they only have minor
consequences~\cite{kwon}.

The Berry phases are more crucial in the paramagnet state, with
full spin rotation symmetry, expected at large $g$. Because this
is a strongly coupled system in the 3 dimensions with a
non-abelian symmetry, no exact duality analysis is possible
(unlike the simpler situation in Section~\ref{xy}). We proceed
with the strategy discussed in Ref.~\cite{kwon}: we replace the
integral over the ${\bf n}_j$ by an integral over the
$\mathcal{A}_{j\mu}$ the variables that appear in the Berry
phases. Formally, this can be done by introducing new `dummy'
variables $A_{j \mu}$ and rewriting (\ref{f1}) by introducing
factors of unity on each link; this leads to
\begin{eqnarray}
Z &=& \int \prod_{j \mu} d A_{j \mu} \exp \left(- i S \sum_j
\eta_j A_{j\tau} \right) \int \prod_j d {\bf n}_j \delta({\bf
n}_j^2 - 1) \delta(\mathcal{A}_{j\mu} - A_{j\mu}) \nonumber \\
&~&~~~~~~~~~~~~~~~~~~~~~~~~~~~~~~~~~~~~~~~~\times\exp \left( -
\frac{1}{2g} \sum_{j,\mu} {\bf n}_j \cdot {\bf n}_{j + \hat{\mu}}
\right)
\nonumber \\
&=& \int \prod_{j \mu} d A_{j \mu} \exp \left(-\mathcal{S}_A
(A_{j\mu}) - i S \sum_j \eta_j A_{j\tau} \right).
 \label{f2a}
\end{eqnarray}
In the first expression, if the integral over the $A_{j \mu}$ is
performed first, we trivially return to (\ref{f1}); however in the
second expression we perform the integral over the ${\bf n}_j$
variables first at the cost of introducing an unknown effective
action $\mathcal{S}_A$ for the $A_{j \mu}$. In principle,
evaluation of $\mathcal{S}_A$ may be performed order-by-order in a
``high temperature'' expansion in $1/g$: we match correlators of
the $A_{j \mu}$ flux with those of the $\mathcal{A}_{j \mu}$ flux
evaluated in the integral over the ${\bf n}_j$ with positive
weights determined only by the $1/g$ term in (\ref{f1}). Rather
than undertaking this laborious calculation, we can guess
essential features of the effective action $\mathcal{S}_A$ from
some general constraints. First, correlations in the ${\bf n}_j$
decay exponentially rapidly for large $g$ (with a correlation
length $\sim 1/\ln(g)$), and so $\mathcal{S}_A$ should be local.
Second, it should be invariant under the lattice form of the gauge
transformation (\ref{gauge})
\begin{equation}
A_{j \mu}^{\prime} = A_{j \mu} - \Delta_{\mu} \phi_j \label{f3}
\end{equation}
associated with the change in the reference point on the unit
sphere from ${\bf N}_0$ to ${\bf N}_0^{\prime}$, with $\phi_j$
equal to the area of the spherical triangle formed by ${\bf n}_j$,
${\bf N}_0$ and ${\bf N}_0^{\prime}$. Finally the area of any
triangle on the sphere is uncertain modulo $4 \pi$ and so the
effective action should be invariant under
\begin{equation}
A_{j \mu} \rightarrow A_{j \mu} + 4 \pi. \label{f4}
\end{equation}
The simplest local action which is invariant under (\ref{f3}) and
(\ref{f4}) is that of {\em compact U(1) quantum electrodynamics}
and so we have
\begin{equation}
Z = \int \prod_{j\mu} d A_{j \mu} \exp \left( \frac{1}{e^2}
\sum_{\Box} \cos\left( \frac{1}{2} \epsilon_{\mu\nu\lambda}
\Delta_{\nu} A_{j \lambda} \right) - i S \sum_j \eta_j A_{j\tau}
\right), \label{f5}
\end{equation}
for large $g$; comparison with the large $g$ expansion shows that
the coupling $e^2 \sim g^2$. As in Section~\ref{xy} our analysis
is aided by replacing the cosine interaction in (\ref{f5}) by a
Villain sum over periodic Gaussians:
\begin{equation}
Z = \sum_{\{q_{\bar{\jmath}\mu}\}} \int \prod_{j\mu} d A_{j \mu}
\exp \left( -\frac{1}{2e^2} \sum_{\Box} \left(
\frac{1}{2}\epsilon_{\mu\nu\lambda} \Delta_{\nu} A_{j \lambda} - 2
\pi q_{\bar{\jmath}\mu} \right)^2 - i S \sum_j \eta_j A_{j\tau}
\right), \label{f6}
\end{equation}
where $q_{\bar{\jmath}\mu}$ are integers on the links of the dual
lattice, which pierce the plaquettes of the direct lattice.

\begin{figure}
\centerline{\includegraphics[width=2.5in]{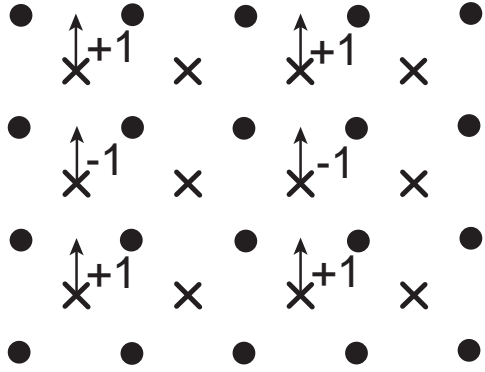}}
\caption{Specification of the non-zero values of the fixed field
$a_{\bar{\jmath}\mu}^0$. The circles are the sites of the direct
lattice, $j$, while the crosses are the sites of the dual lattice,
$\bar{\jmath}$; the latter are also offset by half a lattice
spacing in the direction out of the paper (the $\mu = \tau$
direction). The $a_{\bar{\jmath}\mu}^0$ are all zero for
$\mu=\tau,x$, while the only non-zero values of
$a_{\bar{\jmath}y}^0$ are shown above. Notice that the $a^0$ flux
obeys (\protect\ref{d2}).}\label{fig7}
\end{figure}

We will now perform a series of exact manipulations on (\ref{f6}):
our final result, in (\ref{he1}), will be a demonstration of its
exact equivalence to another height model, but now with the
interface 3=2+1 dimensional. As in Section~\ref{xy}, this
interface model is obtained after a duality mapping and has only
positive weights. This last fact, of course, makes it much more
amenable to a standard statistical analysis. This first step in
the duality transformation is to rewrite (\ref{f6}) by the Poisson
summation formula:
\begin{eqnarray}
\sum_{\{q_{\bar{\jmath}\mu}\}} \exp && \left( -\frac{1}{2e^2}
\sum_{\Box} \left( \frac{1}{2}\epsilon_{\mu\nu\lambda}
\Delta_{\nu} A_{j \lambda} - 2 \pi q_{\bar{\jmath}\mu} \right)^2
\right) \nonumber \\
&&~~~~~~~~~~= \sum_{\{a_{\bar{\jmath}\mu}\}} \exp \left( -
\frac{e^2}{2} \sum_{\bar{\jmath}} a_{\bar{\jmath}\mu}^2 - i
\sum_{\Box} \frac{1}{2} \epsilon_{\mu\nu\lambda}
a_{\bar{\jmath}\mu} \Delta_{\nu} A_{j \lambda}\right), \label{d1}
\end{eqnarray}
where $a_{\bar{\jmath}\mu}$ (like $q_{\bar{\jmath}\mu}$) is an
integer-valued vector field on the links of the dual lattice.
Next, we write the Berry phase in a form more amenable to duality
transformations. Choose a `background' $a_{\bar{\jmath}
\mu}=a_{\bar{\jmath}}^0$ flux which satisfies
\begin{equation}
\epsilon_{\mu\nu\lambda} \Delta_{\nu} a_{\bar{\jmath}\lambda}^0 =
\eta_j \delta_{\mu \tau}, \label{d2}
\end{equation}
where $j$ is the direct lattice site in the center of the
plaquette defined by the curl on the left-hand-side. Any
integer-valued solution of (\ref{d2}) is an acceptable choice for
$a_{\bar{\jmath}\mu}^0$, and a convenient choice is shown in
Fig~\ref{fig7}.
Using (\ref{d2}) to rewrite the Berry phase in (\ref{f6}),
applying (\ref{d1}), and shifting $a_{\bar{\jmath}\mu}$ by the
integer $2S a_{\bar{\jmath}\mu}^0$, we obtain a new exact
representation of $Z$ in (\ref{f6}):
\begin{equation}
Z = \sum_{\{ a_{\bar{\jmath} \mu} \}}  \int \prod_{j\mu} d A_{j
\mu} \exp \left( -\frac{e^2}{2} \sum_{\bar{\jmath},\mu}
(a_{\bar{\jmath}\mu}-2S a_{\bar{\jmath}\mu}^0)^2 - i \sum_{\Box}
\frac{1}{2} \epsilon_{\mu\nu\lambda} a_{\bar{\jmath}\mu}
\Delta_{\nu} A_{j \lambda} \right). \label{d4}
\end{equation}
The integral over the $A_{j \mu}$ can be performed independently
on each link, and its only consequence is the imposition of the
constraint $\epsilon_{\mu\nu\lambda} \Delta_{\nu}
a_{\bar{\jmath}\lambda}=0$. We solve this constraint by writing
$a_{\bar{\jmath} \mu}$ as the gradient of a integer-valued height
$h_{\bar{\jmath}}$ on the sites of the dual lattice, and so obtain
\begin{equation}
Z = \sum_{\{ h_{\bar{\jmath}} \}} \exp \left( -\frac{e^2}{2}
\sum_{\bar{\jmath},\mu} (\Delta_{\mu} h_{\bar{\jmath}}-2S
a_{\bar{\jmath}\mu}^0)^2  \right). \label{d5}
\end{equation}
This is the 2+1 dimensional height model in almost its final form.
Before we proceed to an analysis of (\ref{d5}), we pause for a
paragraph to make some parenthetic remarks on the application of
the above approach to one dimension.

Precisely the same approach as above can be applied to SU(2)
symmetric spin chains which we can model by the obvious
generalization of (\ref{f1}) to 1+1 dimensions. In the large $g$
limit, we proceed with the same set of duality mappings, and in
place of (\ref{d5}) we obtain the partition function
\begin{equation}
Z_1 = \sum_{\{ a_{\bar{\jmath}} \}} \delta_{\Delta_{\mu}
a_{\bar{\jmath}},0} \exp \left( -\frac{e^2}{2} \sum_{\bar{\jmath}}
(a_{\bar{\jmath}}-2S \ell_{\bar{\jmath}\mu})^2 \right).
\label{d51}
\end{equation}
The zero gradient constraint on $a_{\bar{\jmath}}$ (to be
contrasted with the zero curl constraint on $a_{\bar{\jmath}\mu}$
in 2+1 dimensions) means that it takes a single fixed value
through the entire system. For integer $S$, the value
$a_{\bar{\jmath}}=S$ minimizes the action (its action per site is
$e^2 S^2/2$, to be contrasted with the two-fold degenerate BC
states with $a_{\bar{\jmath}}=S-1,S+1$ with action per site $e^2
(S^2+1)/2$), and yields a `featureless' state with no broken
symmetry: this is, of course the Haldane gap state expected for
SU(2) symmetric, integer $S$, spin chains, as discussed in
Section~\ref{int}. For half-odd-integer $S$, there is BC order, as
the action in minimized by the two-fold degenerate values
$a_{\bar{\jmath}}=S+1/2, S-1/2$. All of these results in 1+1
dimensions are in complete accord with the results obtained in
Section~\ref{xy} for large $g$, and with the known properties of
quantum spin chains.

\begin{figure}
\centerline{\includegraphics[width=4in]{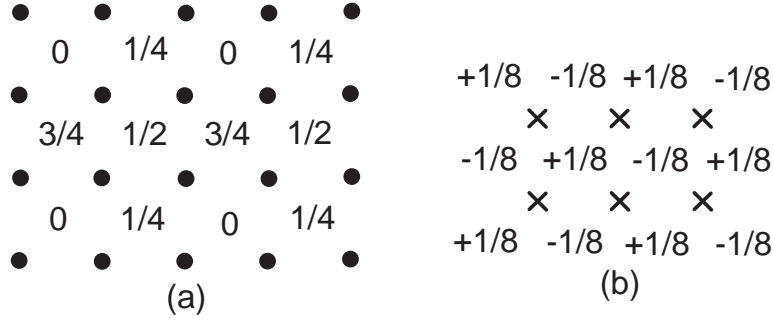}}
\caption{Specification of the non-zero values of the fixed fields
(a) $\mathcal{X}_{\bar{\jmath}}$ and (b) $\mathcal{Y}_{j \mu}$
introduced in (\protect\ref{XY}). The notational conventions are
as in Fig~\protect\ref{fig7}. Only the $\mu=\tau$ components of
$\mathcal{Y}_{j \mu}$ are non-zero, and these are shown in (b).
}\label{fig8}
\end{figure}

Now we return to the 2+1 dimensional case of interest in this
section, and proceed with our interrupted analysis of (\ref{d5}).
We can convert the ``frustration'' $a_{\bar{\jmath}\mu}^0$ in
(\ref{d5}) into offsets for the allowed height values (as in
Section~\ref{xy}) by decomposing it into curl and divergence free
parts and writing it in terms of new fixed fields,
$\mathcal{X}_{\bar{\jmath}}$ and ${\mathcal Y}_{j \mu}$ as
follows:
\begin{equation}
a_{\bar{\jmath}\mu}^{0} = \Delta_{\mu} \mathcal{X}_{\bar{\jmath}}
+ \epsilon_{\mu\nu\lambda} \Delta_{\nu} \mathcal{Y}_{j  \lambda}.
\label{XY}
\end{equation}
The values of these new fields are shown in Fig~\ref{fig8}.
Inserting (\ref{XY}) into (\ref{d5}), we can now write the height
model in its simplest form~\cite{rs1,rs2,fradkiv}
\begin{equation}
Z_h = \sum_{\{H_{\bar{\jmath}}\}} \exp \left ( - \frac{e^2}{2}
\sum_{\bar{\jmath}} \left( \Delta_{\mu} H_{\bar{\jmath}} \right)^2
\right), \label{he1}
\end{equation}
where
\begin{equation}
H_{\bar{\jmath}} \equiv h_{\bar{\jmath}} - 2 S
\mathcal{X}_{\bar{\jmath}} \label{he2}
\end{equation}
is the new height variable we shall work with. Notice that the
$\mathcal{Y}_{j \mu}$ have dropped out, while the
$\mathcal{X}_{\bar{\jmath}}$ act only as fractional offsets (for
$S$ not an even integer) to the integer heights. From (\ref{he2})
we see that for half-odd-integer $S$ the height is restricted to
be an integer on one of the four sublattices, an integer plus 1/4
on the second, an integer plus 1/2 on the third, and an integer
plus 3/4 on the fourth; the fractional parts of these heights are
as shown in Fig~\ref{fig8}a; the steps between neighboring heights
are always an integer plus 1/4, or an integer plus 3/4. For $S$ an
odd integer, the heights are integers on one square sublattice,
and half-odd-integers on the second sublattice. Finally for even
integer $S$ the offset has no effect and the height is an integer
on all sites. We discuss these classes of $S$ values in turn in
the following subsections.

(It is worth emphasizing here that the height model $Z_h$ in
(\ref{he1}) only offers a description of the large $g$
paramagnetic phase of the model $Z$ in (\ref{f1}) in 2+1
dimensions. The small $g$ N\'{e}el ordered state is not one of the
phases of $Z_h$, nor is the critical point describing the onset of
magnetic order; an extension of $Z_h$ to include the N\'{e}el
state was discussed in~\cite{kwon}, but will not be presented
here. This should be contrasted with the situation in 1+1
dimensions discussed in Section~\ref{xy}; there height models like
$Z_{XY}$ in (\ref{e10}) can describe all the phases and critical
points of the quantum spin chain, including the TL phase, which is
a state with quasi-long-range N\'{e}el order).

\subsection{$S$ even integer}
\label{S2}

Unlike the two-dimensional case, three-dimensional height models
generically have no roughening transition, and the interface is
always smooth. With all heights integers, the smooth phase breaks
no symmetries. So like the case of integer $S$ in quantum spin
chains (Section~\ref{int}), square lattice antiferromagnets with
$S$ even integer can have a paramagnetic ground state with a spin
gap and no broken symmetries. This is in accord with the exact
ground state for a $S=2$ antiferromagnet on the square lattice
found by Affleck {\em et al.}~\cite{aklt}, the AKLT state. The
structure of this state is identical to that of the paramagnetic
state of the $\varphi^4$ field theory in 2+1 dimensions (Eqn
(\ref{zl1}) with $c_x=c_y$), with a stable $S=1$ spin exciton
excitation. So the critical point between the magnetic N\'{e}el
state and the spin gap state should also be described by the
$\varphi^4$ field theory. As we noted above, this critical point
is not contained within $Z_h$.

\subsection{$S$ half-odd-integer}
\label{Shint}

Now the heights of the interface model can take four possible
values, which are integers plus the offsets on the four square
sublattices shown in Fig~\ref{fig8}a. As in Section~\ref{S2}, the
interface is always smooth, but as in Section~\ref{hint}, any
smooth interface must break a lattice symmetry with the
development of bond-centered charge (BC) order and this allows a
number of distinct spin gap ground states of the lattice
antiferromagnet. Consistent with these theoretical predictions,
numerical studies of the frustrated spin model (\ref{f2}) have
produced evidence for BC order in the state contiguous to the
N\'{e}el state: see~\cite{sushkov,leiden} for recent results.

\begin{figure}
\centerline{\includegraphics[width=2.3in]{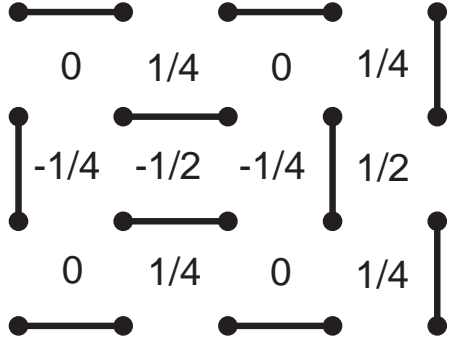}}
\caption{Mapping between the quantum dimer model and the interface
model $Z$ in (\protect\ref{he1}). Each dimer on the direct lattice
is associated with a step in height of $\pm 3/4$ on the link of
the dual lattice that crosses it. All other height steps are $\pm
1/4$. Each dimer represents a singlet valence between the sites,
as in Fig~\protect\ref{fig2}.}\label{fig9}
\end{figure}

It is useful, first, to obtain a simple physical interpretation of
the interface model in the language of the $S=1/2$
antiferromagnet~\cite{zheng}. From Fig~\ref{fig8}a it is clear
that nearest neighbor heights can differ either by 1/4 or 3/4
(modulo integers). To minimize the action in (\ref{he1}), we
should choose the interface with the largest possible number of
steps of $\pm 1/4$. However, the interface is frustrated, and it
is not possible to make all steps $\pm 1/4$ and at least a quarter
of the steps must be $\pm 3/4$. Indeed, there is a precise
one-to-one mapping between interfaces with the minimal number of
$\pm 3/4$ steps (we regard interfaces differing by a uniform
integer shift in all heights as equivalent) and the dimer
coverings of the square lattice: the proof of this claim is
illustrated in Fig~\ref{fig9}.
We identify each dimer with a singlet valence bond between the
spins (the ellipses in Fig~\ref{fig2}), and so each interface
corresponds to a quantum state with each spin locked in the a
singlet valence bond with a particular nearest neighbor.
Fluctuations of the interface in imaginary time between such
configurations correspond to quantum tunneling events between such
dimer states, and an effective Hamiltonian for this is provided by
the quantum dimer model~\cite{qd1,qd2}.

\begin{figure}
\centerline{\includegraphics[width=2.7in]{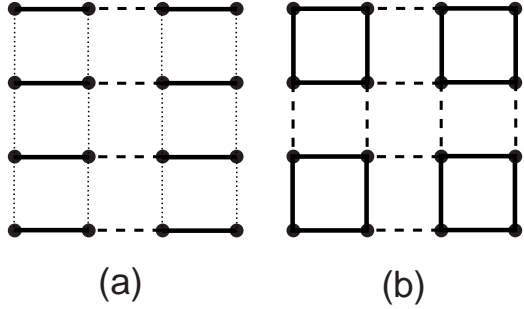}}
\caption{Sketch of the two simplest possible states with
bond-centered charge (BC) order for $S=1/2$ on the square lattice:
(a) the columnar spin-Peierls states, and (b) plaquette state. The
different values of the $\langle {\bf S}_i \cdot {\bf S}_j
\rangle$ on the links are encoded by the different line styles.
Both states are 4-fold degenerate; an 8-fold degenerate state,
with superposition of the above orders, also appears as a possible
ground state of the generalized interface model.} \label{fig10}
\end{figure}

The nature of the possible smooth phases of the interface model
are easy to determine from the above picture, and by standard
techniques from statistical theory which have been reviewed
elsewhere~\cite{rs2,kwon,zheng,sr}. Interfaces with average height
$\langle H_{\bar{\jmath}} \rangle = 1/8,3/8,5/8,7/8$ (modulo
integers) correspond to the four-fold degenerate BC ordered states
in Fig~\ref{fig10}a, while those with $\langle H_{\bar{\jmath}}
\rangle = 0,1/4,1/2,3/4$ (modulo integers) correspond to the
four-fold degenerate plaquette BC states in Fig~\ref{fig10}b.
All other values of $\langle H_{\bar{\jmath}} \rangle$ are
associated with eight-fold degenerate BC states with a
superposition of the orders in Fig~\ref{fig10}a and b.

 It is interesting to note
here that the symmetry of the state in Fig~\ref{fig10}a is
identical to the symmetry of the coupled ladder Hamiltonian
illustrated in Fig~\ref{fig1}: so the system has spontaneously
decided to arrange itself in the form of a coupled ladder. Whether
the BC ordering is spontaneous (as it is here) or explicitly
present in the Hamiltonian (as in Section~\ref{ladders}), the
nature of the non-zero spin excitations should be the same. So the
confinement argument of Fig~\ref{fig3}b applies here too, and
there are no free $S=1/2$ spinons in the BC state. The elementary
excitation is a $S=1$ spin exciton. This should be contrasted with
the $S=1/2$ quantum spin chain in Section~\ref{hint}, where the BC
state had spinon excitations.

We are now finally in a position to explain a central step in the
analysis of Section~\ref{ladders}: the neglect of Berry phases in
the theory of the magnetic critical point for the coupled ladder
antiferromagnet. We have seen here that the primary effect of the
Berry phases is to induce BC order in the spin gap state. However,
such ``order'' is already present explicitly in the Hamiltonian of
the coupled ladder antiferromagnet, which has two sites per unit
cell: the Hamiltonian picks out one of the four BC states and
``dimer'' fluctuations to the other BC states are suppressed. So
we need only focus on the fluctuations of the magnetic order, and
these are captured by the $\varphi^4$ field theory.

The nature of the crossover between the N\'{e}el and BC states for
$S=1/2$ quantum spin models with the full square lattice symmetry
(like $H$ in (\ref{f2})) is far more subtle, and was recently
discussed at some length in Ref~\cite{kwon}. Here, both the
N\'{e}el and BC orders can become critical, and so we cannot
naively assert the irrelevance of the Berry phases. One appealing
scenario~\cite{sushkov,kwon}, supported by recent numerical
studies~\cite{sushkov}, is that the crossover occurs via two
transitions: with increasing $g$, there is first a transition at
$g=g_{c1}$ to a state with co-existing N\'{e}el and BC order (only
the BC fluctuations are critical at $g_{c1}$ are these are
described by a $Z_4$ clock model), and then a transition at
$g=g_{c2} > g_{c1}$ where N\'{e}el order disappears. The second
transition at $g_{c2}$ is just as in the coupled ladder
antiferromagnet, and is therefore described by the $\varphi^4$
field theory: there is well-formed BC order on the either side of
$g=g_{c2}$, and it is immaterial here whether this order was
explicitly present in the Hamiltonian or spontaneously induced.
Indeed, this was secretly the reason we chose to consider the
coupled ladder antiferromagnet in our early discussion.

\subsection{$S$ odd integer}

This case is similar to that $S$ half-odd-integer, and we will not
consider it in detail. The Berry phases again induce BC order in
the spin gap state, but this order need only lead to a two-fold
degeneracy.

\section{Magnetic transitions in $d$-wave superconductors}
\label{magd}

We have so far considered a variety of magnetic transitions in
correlated ``Mott'' insulators, which have an electron density
commensurate with the number of available orbitals. Cuprate high
temperature superconductivity appears when such a Mott insulator
on the square lattice is doped by a finite density of mobile
carriers. All of the available experimental evidence supports the
proposal that, apart from some inhomogeneous charge ordered
insulating states, the ground state of the doped Mott insulator is
a superconductor. In particular, there is no clear-cut evidence
for a metallic Fermi liquid state up to a moderate density of
holes. Consequently, we will focus our attention on the
superconducting ground state in the remainder of this paper. This
is to be contrasted with the approach of~\cite{rome} which
examines transitions in metallic states at non-zero temperature.

It is now natural to study the interplay of the N\'{e}el magnetic
order found in the insulator and the superconducting order of the
doped system. (The charge order discussed in Section~\ref{2d} also
plays an important role in the superconductor, but we will not
focus attention on this aspect here --- see~\cite{sns} for a
recent review.) The experimental evidence indicates that this
crossover proceeds via an intermediate state with {\em
co-existence} of magnetic and superconducting order. Furthermore,
the magnetic order is no longer in the simple two-sublattice
N\'{e}el configuration, but oscillates at a wavevector $K \neq
(\pi,\pi)$. So instead of expressing the electron spin as
$S_{j\alpha} = \eta_j \varphi_{\alpha} (r_j)$, which is implied by
(\ref{cont}) and the subsequent coarse-graining of ${\bf n}$ to
$\varphi_{\alpha}$, we now write
\begin{equation}
S_{j \alpha} = \Phi_{\alpha} (r_j ) e^{i K r} + \mbox{c.c.},
\label{g1}
\end{equation}
where $\Phi_{\alpha} (r,\tau)$ a 3-component, {\em complex}
quantum field; $\Phi_{\alpha}$ is required to be complex as along
as the magnetic state does not have a two sublattice structure,
which is only the case for ordering wavevectors $(\pi,\pi)$ and
$(0,\pi)$. Also, the experimental evidence also indicates that the
magnetic order is {\em collinear}, and not spiral; this is the
case if $\epsilon_{\alpha\beta\gamma} \langle \Phi_{\beta} \rangle
\langle \Phi_{\gamma}^{\ast} \rangle = 0$.

With this reasoning, we are led to consider the quantum transition
between a superconducting state with co-existing magnetic
spin-density-wave order at wavevector $K$ ($\langle \Phi_{\alpha}
\rangle \neq 0$ in the SC+SDW state) to an ordinary superconductor
($\langle \Phi_{\alpha} \rangle = 0$ in the SC state). Clearly,
the field $\Phi_{\alpha}$ should be a degree of freedom in any
theory for this transition. However, in a superconductor there are
additional fermionic excitations which carry spin $S=1/2$, the
Bogoliubov quasiparticles, and we have to examine their influence
on the critical properties. This is especially the case in a
$d$-wave superconductor, in which the energy of the quasiparticles
vanishes at four `nodal' points in the Brillouin zone.

To address this issue, let us review the standard BCS mean-field
theory for a $d$-wave superconductor on the square lattice. We
consider the Hamiltonian
\begin{equation}
H_{tJ} = \sum_{k} \varepsilon_k c_{k \sigma}^{\dagger} c_{k
\sigma} + J_1 \sum_{\langle ij \rangle} {\bf S}_i \cdot {\bf
S}_{j} \label{g2}
\end{equation}
where $c_{j\sigma}$ is the annihilation operator for an electron
on site $j$ with spin $\sigma=\uparrow,\downarrow$, $c_{k\sigma}$
is its Fourier transform to momentum space, $\varepsilon_k$ is the
dispersion of the electrons (it is conventional to choose
$\varepsilon_k = -2t_1 (\cos(k_x) + \cos(k_y)) - 2 t_2 ( \cos(k_x
+ k_y) + \cos(k_x - k_y)) - \mu$, with $t_{1,2}$ the first/second
neighbor hopping and $\mu$ the chemical potential), and the $J_1$
term is the same as that in (\ref{f2}) with
\begin{equation}
S_{j\alpha} = \frac{1}{2} c^{\dagger}_{j\sigma}
\sigma_{\sigma\sigma'}^{\alpha} c_{j \sigma'} \label{g2a}
\end{equation}
and $\sigma^{\alpha}$ the Pauli matrices. We will consider
the consequences of the $J_2$ term in (\ref{f2}) for the
superconductor in Section~\ref{did} below. Applying the BCS
mean-field decoupling to $H_{tJ}$ we obtain the Bogoliubov
Hamiltonian
\begin{equation}
H_{BCS} = \sum_{k} \varepsilon_k c_{k \sigma}^{\dagger} c_{k
\sigma} - \frac{J_1}{2} \sum_{j\mu}\Delta_{\mu} \left(
c^{\dagger}_{j\uparrow} c^{\dagger}_{j+\hat{\mu},\downarrow} -
c^{\dagger}_{j\downarrow} c^{\dagger}_{j+\hat{\mu},\uparrow}
\right) + \mbox{h.c.}. \label{g3}
\end{equation}
For a wide range of parameters, the ground state energy optimized
by a $d_{x^2-y^2}$ wavefunction for the Cooper pairs: this
corresponds to the choice $\Delta_x = - \Delta_y =
\Delta_{x^2-y^2}$. The value of $\Delta_{x^2-y^2}$ is determined
by minimizing the energy of the BCS state
\begin{equation}
E_{BCS} = J_1 |\Delta_{x^2-y^2}|^2 - \int \frac{d^2 k}{4 \pi^2}
\left[ E_k - \varepsilon_k \right] \label{g4}
\end{equation}
where the fermionic quasiparticle dispersion is
\begin{equation}
E_k = \left[ \varepsilon_k^2 + \left|J_1 \Delta_{x^2-y^2}(\cos k_x
- \cos k_y)\right|^2 \right]^{1/2}. \label{g5}
\end{equation}

The energy of the quasiparticles, $E_k$, vanishes at the four
points $(\pm Q, \pm Q)$ at which $\varepsilon_k=0$. We are
especially interested in the low energy quasiparticles in the
vicinity of these points, and so we perform a gradient expansion
of $H_{BCS}$ near each of them. We label the points $Q_1=(Q,Q)$,
$Q_2=(-Q,Q)$, $Q_3=(-Q,-Q)$, $Q_4=(Q,-Q)$ and write
\begin{equation}
c_{j\sigma} =f_{1\sigma} (x_j) e^{i Q_1 x_j}+f_{2\sigma} (x_j)
e^{i Q_2 x_j}+f_{3\sigma} (x_j) e^{i Q_3 x_j}+f_{4\sigma} (x_j)
e^{i Q_4 x_j},\label{g5a}
\end{equation}
while assuming the $f_{1-4,\sigma} (x)$ are slowly varying
functions of $x$. We also introduce the bispinors $\Psi_1 =
(f_{1\uparrow}, f_{3\downarrow}^{\dagger},
f_{1\downarrow},-f_{3\uparrow}^{\dagger})$, and $\Psi_2 =
(f_{2\uparrow}, f_{4\downarrow}^{\dagger},
f_{2\downarrow},-f_{4\uparrow}^{\dagger})$, and then express
$H_{BCS}$ in terms of $\Psi_{1,2}$ while performing a spatial
gradient expansion. This yields the following effective action for
the fermionic quasiparticles:
\begin{eqnarray}
\mathcal{S}_{\Psi} = \int d \tau d^2 x && \left[ \Psi_1^{\dagger}
\left(
\partial_{\tau} - i v_F \tau^z \partial_x - i v_{\Delta} \tau^x
\partial_y \right)\Psi_1 \right. \nonumber \\
&&~~~~~\left. + \Psi_2^{\dagger} \left(
\partial_{\tau} - i v_F \tau^z \partial_y - i v_{\Delta} \tau^x
\partial_x \right)\Psi_2 \right],
\label{g6}
\end{eqnarray}
where we have rotated our spatial co-ordinate system by
$45^\circ$, and the $\tau^{x,z}$ are $4 \times 4$ matrices which
are block diagonal, the blocks consisting of $2\times 2$ Pauli
matrices. The velocities $v_{F,\Delta}$ are given by the conical
structure of $E_k$ near the $Q_{1-4}$: we have $v_F =
\left|\nabla_k \varepsilon_k |_{k=Q_a} \right|$ and $v_{\Delta} =
|J_1 \Delta_{x^2-y^2} \sqrt{2} \sin (Q)|$.

We are now in a position to discuss the quantum field theory of
the transition from the SC+SDW to the SC state in term of the
$\Phi_{\alpha}$ and $\Psi_{1,2}$ degrees of freedom. First, we can
write down an action for $\Phi_{\alpha}$ along the lines of that
for $\varphi_{\alpha}$ in the insulator: this will have a
structure very similar to that of (\ref{zl1}) (the complex nature
of $\Phi_{\alpha}$ induces additional quartic terms which we will
not discuss here---the reader is referred to another review by the
author~\cite{sns} for details on this point). The action will also
contain the terms in $\mathcal{S}_{\Psi}$ in (\ref{g6}), and it
remains to discuss terms which may couple the $\Phi_{\alpha}$ and
$\Psi_{1,2}$. The simplest possible terms are cubic interaction
terms like $\Phi_{\alpha} \Psi_{1} \Psi_1$ etc. However by
comparing the momentum dependencies in (\ref{g1}) and (\ref{g5a})
with (\ref{g2a}) it is easy to see that such terms will survive
after averaging over space only if the special commensurability
conditions $K = (2Q,2Q)$, or $K=(2Q,0)$, are satisfied. In general
this will not be the case, and experimental evidence also does not
support this possibility. Therefore, we will assume in the
remainder of this section that these commensurability conditions
are not satisfied: the theory for the case $K=(2Q,2Q)$, $Q=\pi/2$
was considered in Ref.~\cite{bfn}, and has some formal
similarities to the theory to be considered in Section~\ref{did}
in a different physical context.

In the absence of special commensurability conditions, the
simplest term which can couple the $\Phi_{\alpha}$ and
$\Psi_{1,2}$ is
\begin{equation}
\kappa \int d^2 r d \tau |\Phi_{\alpha}|^2 \left( \Psi_1^{\dagger}
\Psi_1 + \Psi_2^{\dagger} \Psi_2 \right). \label{g7}
\end{equation}
However a simple scaling argument shows that the coupling $\kappa$
is irrelevant at the SC+SDW to SC critical point. To see this,
imagine that the critical point is described by the $\kappa=0$
point, where the $\Phi_{\alpha}$ and $\Psi_{1,2}$ fields are
decoupled. This point is actually a fixed point of the
renormalization group and is invariant under scaling
transformations. From (\ref{g6}) we see that the scaling
dimensions of $\Psi_{1,2}$ is 1, while that of $|\Phi_{\alpha}|^2$
is $(3-1/\nu)$, where $\nu$ is the correlation length exponent. So
from (\ref{g7}) we conclude that the scaling dimension of $\kappa$
is $(1/\nu-2)$. The exponent $\nu$ is determined by a theory like
that in (\ref{zl1}) for $\Psi_{\alpha}$ and all such theories have
$\nu > 1/2$, the mean-field exponent. Consequently the scaling
dimension of $\kappa$ is negative, and the decoupled $\kappa=0$
fixed point describes the SC+SDW to SC quantum phase transition.
The critical magnetic fluctuations are associated with
$\Phi_{\alpha}$, and the fermionic, $S=1/2$ quasiparticles
$\Psi_{1,2}$ are merely innocent spectators.

This last result is quite powerful, as it implies that we can
apply the results of the quantum transition in the insulator,
discussed in Section~\ref{ladders}, essentially unchanged to the
magnetic fluctuations in the cuprate high temperature
superconductors. Such a proposal was first made in
Ref.~\cite{CSY1,CSY2}, and leads to the prediction that the $S=1$
exciton should survive as a stable excitation in the SC state near
its transition to the SC+SDW state, and that this should be
evident as a sharp ``resonance'' in dynamic neutron scattering:
such a resonance appears to have been observed in a number of
experiments~\cite{res1,res2,res3,res4,res5}. The framework of the
SC+SDW to SC transition has also been used to understand dynamic
spin correlations in the presence of Zn
impurities~\cite{imp1,imp2}, and to explore the phase diagram in
the presence of an applied magnetic field~\cite{dsz,pphmf}: the
results have been quantitatively compared with neutron
scattering~\cite{keimer1,keimer2,younglee,lake,lake2} and scanning
tunnelling microscopy experiments~\cite{seamus}.

\section{Transition between BCS superconductors}
\label{did}

In this final section, we will consider one of the simplest
examples of a non-trivial quantum critical point in two spatial
dimensions which, surprisingly, was only discovered
recently~\cite{vojta1,vojta2}. The two phases on either side of
the critical point are described by an entirely conventional BCS
theory: they differ only in the symmetry of the Cooper pair
wavefunction, dependent upon the relative co-ordinates of the
paired electrons. At $T=0$, BCS theory provides an accurate
description of the low energy properties everywhere except at the
point which separates the two phases: this is probably the only
known example of the failure of BCS theory in two (or higher)
dimensions in a superconducting ground state. At $T>0$, this
failure broadens into the ``quantum critical'' region which should
be easily experimentally detectable: we will not discuss these
$T>0$ issues here, and instead refer the reader to other
reviews~\cite{book}.

Our original motivation for examining transitions between BCS
superconductors was provided by photoemission
experiments~\cite{valla} which observed an anomalous broadening of
the quasiparticle spectrum. We will discuss this further in
Section~\ref{expts}, but will motivate the problem now on
theoretical grounds. The Hubbard model on a square lattice is
believed to provide a good starting point for studying the physics
of the cuprate superconductors. There is now good evidence that
for electron densities close to half-filling, the primary
instability of the Fermi liquid state is to a BCS superconductor
with $d_{x^2-y^2}$ symmetry~\cite{metzner}, whose mean-field
theory was presented in Section~\ref{magd}. The opposite limit of
a low density of electrons, well away from half-filling, has also
been examined theoretically~\cite{maxim1,maxim2}, and the low
density expansion permits reliable predictions to be made: it was
found that the ground state was again a BCS superconductor, but
now the pairs had a distinct $d_{xy}$ symmetry (in the presence of
a square lattice, the $d_{x^2-y^2}$ and $d_{xy}$ states are not
degenerate, as they would be in free space). Here we will answer
the simple question: how does the ground state evolve from between
these limits with qualitatively different wavefunctions ?

Rather than working the full complexities of the Hubbard model, we
follow the strategy of Section~\ref{magd} and work with the
simplest phenomenological model that contains the two phases we
are interested in. We want to extend $H_{tJ}$ in (\ref{g2}) so
that a BCS mean-field theory will permit a region with $d_{xy}$
superconductivity. Clearly this will be possible if we have a
pairing interaction along the $(1,\pm 1)$ directions of the square
lattice; so we extend $H_{tJ}$ to allow for a second-neighbor
exchange along the diagonals, $J_2$, as in (\ref{f2}):
\begin{equation}
\widetilde{H}_{tJ} = \sum_{k} \varepsilon_k c_{k \sigma}^{\dagger}
c_{k \sigma} + J_1 \sum_{\langle ij \rangle} {\bf S}_i \cdot {\bf
S}_{j} + J_2 \sum_{{\rm nnn}~ij} {\bf S}_i \cdot {\bf S}_j.
\label{g8}
\end{equation}
We will follow the evolution of the ground state of
$\widetilde{H}_{tJ}$ as a function of $J_2 / J_1$. We expect that
the universal properties of this route will be similar to that of
the Hubbard model as a function of carrier concentration, in light
of the studies of the latter model quoted above.

\subsection{BCS Theory}
\label{bcs}

\begin{figure}
\centerline{\includegraphics[width=2.5in]{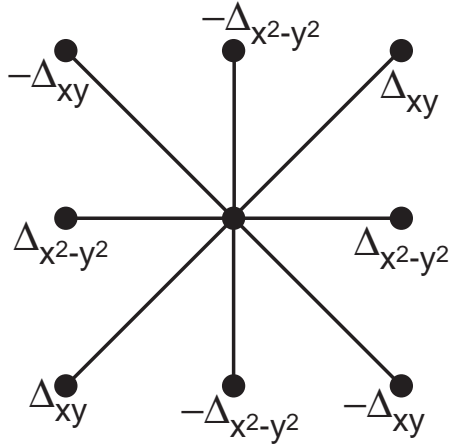}}
\caption{Values of the pairing amplitudes, $-\langle c_{i
\uparrow} c_{j \downarrow} -  c_{i \downarrow} c_{j \uparrow}
\rangle$ with $i$ the central site, and $j$ is one of its 8 near
neighbors.} \label{fig11}
\end{figure}

We begin with a standard BCS analysis of $\widetilde{H}_{tJ}$
which closely parallels that presented in Section~\ref{magd}. The
mean-field Hamiltonian is now modified from (\ref{g3}) to
\begin{eqnarray}
\widetilde{H}_{BCS} = \sum_{k} \varepsilon_k c_{k
\sigma}^{\dagger} c_{k \sigma} &-& \frac{J_1}{2} \sum_{j,\mu}
\Delta_{\mu} (c_{j\uparrow}^{\dagger}
c_{j+\hat{\mu},\downarrow}^{\dagger} - c_{j\downarrow}^{\dagger}
c_{j+\hat{\mu},\uparrow}^{\dagger}) + \mbox{h.c.} \nonumber \\ &-&
\frac{J_2}{2} {\sum_{j,\nu}}^{\prime} \Delta_{\nu}
(c_{j\uparrow}^{\dagger} c_{j+\hat{\nu},\downarrow}^{\dagger} -
c_{j\downarrow}^{\dagger} c_{j+\hat{\nu},\uparrow}^{\dagger}) +
\mbox{h.c.}, \label{g9}
\end{eqnarray}
where the second summation over $\nu$ is along the diagonal
neighbors $\hat{x}+\hat{y}$ and $-\hat{x}+\hat{y}$. To obtain
$d_{xy}$ pairing along the diagonals, we choose $\Delta_{x+y} = -
\Delta_{-x+y} = \Delta_{xy}$. We summarize our choices for the
spatial structure of the pairing amplitudes (which determine the
Cooper pair wavefunction) in Fig~\ref{fig11}.
The values of $\Delta_{x^2-y^2}$ and $\Delta_{xy}$ are to be
determined by minimizing the ground state energy (generalizing
(\ref{g4}))
\begin{equation}
E_{BCS} = J_1 |\Delta_{x^2-y^2}|^2 +J_2 |\Delta_{xy}|^2 - \int
\frac{d^2 k}{4 \pi^2} \left[ E_k - \varepsilon_k \right]
\label{g10}
\end{equation}
where the quasiparticle dispersion is now (generalizing
(\ref{g5}))
\begin{equation}
E_k = \left[ \varepsilon_k^2 + \left|J_1 \Delta_{x^2-y^2}(\cos k_x
- \cos k_y) + 2 J_2 \Delta_{xy} \sin k_x \sin k_y \right|^2
\right]^{1/2}. \label{g11}
\end{equation}
Notice that the energy depends upon the relative phase of
$\Delta_{x^2-y^2}$ and $\Delta_{xy}$: this phase is therefore an
observable property of the ground state.

\begin{figure}
\centerline{\includegraphics[width=4.5in]{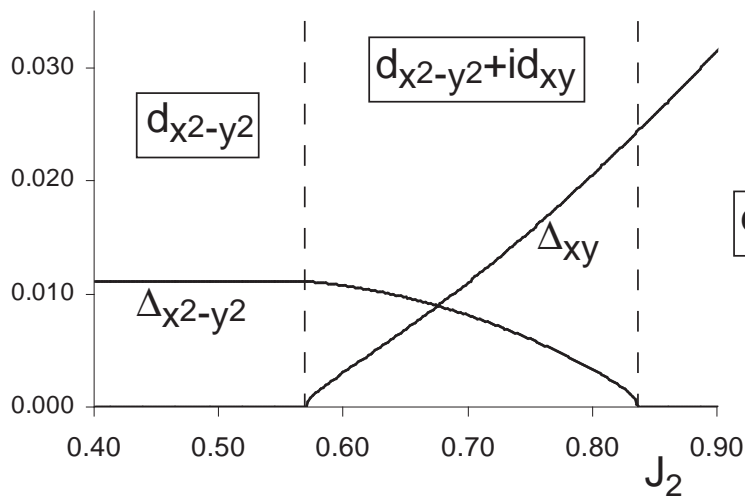}} \caption{BCS
solution of the phenomenological Hamiltonian $\widetilde{H}_{tJ}$
in (\protect\ref{g8}). Shown are the optimum values of the pairing
amplitudes $|\Delta_{x^2-y^2}|$ and $|\Delta_{xy}|$ as a function
of $J_2$ for $t_1 =1$, $t_2=-0.25$, $\mu=-1.25$, and $J_1$ fixed
at $J_1=0.4$. The relative phase of the pairing amplitudes was
always found to obey (\protect\ref{g12}). The dashed lines denote
locations of phase transitions between $d_{x^2-y^2}$,
$d_{x^2-y^2}+id_{xy}$, and $d_{xy}$ superconductors. The pairing
amplitudes vanish linearly at the two critical points,
corresponding to the exponent $\beta_{BCS}=1$ in
(\protect\ref{g14}); the slight rounding at the end points
apparent in the figure is a consequence of the finite step-size
used in numerical evaluation of the integral in
(\protect\ref{g10}). Combining these results with those
of~\protect\cite{metzner,maxim1,maxim2} we propose that the more
realistic Hubbard model will show the same sequence of phases as
those above as a function of increasing doping. Dagan and
Deutscher~\protect\cite{dagan} were the first to propose that
$d_{x^2-y^2}+id_{xy}$ superconductivity occurs at a doping larger
than $d_{x^2-y^2}$ superconductivity on the basis of experiments
we will discuss in Section~\protect\ref{expts}.} \label{fig12}
\end{figure}

It is a simple matter to numerically carry out the minimization of
(\ref{g11}), and the results for a typical choice of parameters
are shown in Fig~\ref{fig12} as a function $J_2/J_1$.
One of the two amplitudes $\Delta_{x^2-y^2}$ or $\Delta_{xy}$ is
always non-zero and so the ground state is always superconducting.
The transition from pure $d_{x^2-y^2}$ superconductivity to pure
$d_{xy}$ superconductivity occurs via an intermediate phase in
which {\em both} order parameters are non-zero. Furthermore, in
this regime, their relative phase is found to be pinned to $\pm
\pi/2$ {\em i.e.}
\begin{equation}
\arg (\Delta_{xy}) = \arg (\Delta_{x^2-y^2}) \pm \pi/2 \label{g12}
\end{equation}
The reason for this pinning can be intuitively seen from
(\ref{g11}): only for these values of the relative phase does the
equation $E_k = 0$ never have a solution. In other words, the
gapless nodal quasiparticles of the $d_{x^2-y^2}$ superconductor
(and also those of the $d_{xy}$ superconductor) acquire a finite
energy gap when a secondary pairing with relative phase $\pm
\pi/2$ develops. By a level repulsion picture, we can expect that
gapping out the low energy excitations should help lower the
energy of the ground state. The intermediate phase obeying
(\ref{g12}) is called a $d_{x^2-y^2} + i d_{xy}$ superconductor.

The choice of the sign in (\ref{g12}) leads to an overall two-fold
degeneracy in the choice of the wavefunction for the $d_{x^2-y^2}
+ i d_{xy}$ superconductor. This choice is related to the breaking
of time-reversal symmetry, and implies that the $d_{x^2-y^2} + i
d_{xy}$ phase is characterized by the non-zero expectation value
of a $Z_2$ Ising order parameter; the expectation value of this
order vanishes in the two phases (the $d_{x^2-y^2}$ and $d_{xy}$
superconductors) on either side of the $d_{x^2-y^2}+id_{xy}$
superconductor. As is conventional, we will represent the Ising
order by a real scalar field $\phi$. Fluctuations of $\phi$ become
critical near both of the phase boundaries in Fig~\ref{fig12}. As
we will explain in Section~\ref{qft} below, the critical theory
for $\phi$ fluctuations is {\em not} the usual $\phi^4$ field
theory which describes the ordinary Ising transition in three
spacetime dimensions.

Near the phase boundary from $d_{x^2-y^2}$ to $d_{x^2-y^2} +
id_{xy}$ superconductivity it is clear that we can identify
\begin{equation}
\phi = i \Delta_{xy}, \label{o1}
\end{equation}
(in the gauge where $\Delta_{x^2-y^2}$ is real). We can now expand
$E_{BCS}$ in (\ref{g10}) for small $\phi$ (with $\Delta_{x^2-y^2}$
finite) and find a series with the structure~\cite{laugh,wolf}
\begin{equation}
E_{BCS} = E_0 + s \phi^2 + v |\phi|^3 + \ldots, \label{g13}
\end{equation}
where $s$, $v$ are coefficients and the ellipses represent regular
higher order terms in even powers of $\phi$; $s$ can have either
sign, whereas $v$ is always positive. Notice the non-analytic
$|\phi|^3$ term that appears in the BCS theory
--- this arises from an infrared singularity in the integral in
(\ref{g10}) over $E_k$ at the four nodal points of the
$d_{x^2-y^2}$ superconductor, and is a preliminary indication that
the transition differs from that in the ordinary Ising model. We
will achieve a deeper understanding of this non-analyticity when
we consider fluctuations in Section~\ref{qft}. We can optimize
$\phi$ by minimizing $E_{BCS}$ in (\ref{g13})--- this shows that
$\langle \phi \rangle =0$ for $s>0$, and $\langle \phi \rangle
\neq 0$ for $s<0$. So $s \sim (J_2/J_1)_c - J_2/J_1$ where
$(J_2/J_1)_c$ is the first critical value in Fig~\ref{fig12}. Near
this critical point, we find
\begin{equation}
\langle \phi \rangle \sim (s_c - s)^{\beta}, \label{g14}
\end{equation}
where we have allowed for the fact that fluctuation corrections
will shift the critical point from $s=0$ to $s=s_c$. The present
BCS theory yields the exponent $\beta_{BCS} = 1$; this differs
from the usual mean-field exponent $\beta_{MF} = 1/2$, and this is
of course due to the non-analytic $|\phi|^3$ term in (\ref{g13}).

An essentially identical structure appears at the second critical
point in Fig~\ref{fig12} at the boundary between the
$d_{x^2-y^2}+i d_{xy}$ and $d_{xy}$ superconductors. The
$d_{x^2-y^2}$ and $d_{xy}$ pairing amplitudes exchange roles, and
the Ising order parameter for the transition is $\phi = i
\Delta_{x^2-y^2}$ (replacing (\ref{o1})). We will therefore not
discuss this case further.

\begin{figure}
\centerline{\includegraphics[width=5in]{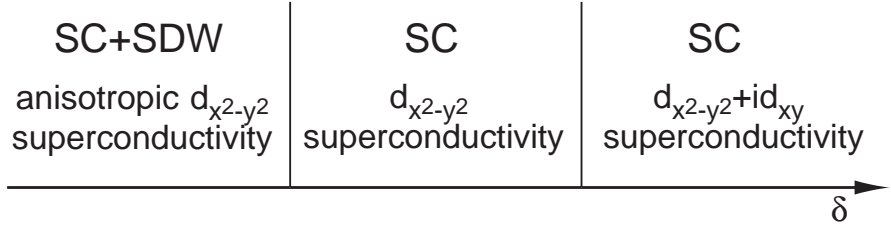}}
\caption{Conjectured $T=0$ phase diagram for the cuprate
superconductors as a function of increasing hole concentration
$\delta$. The theory for the SC+SDW to SDW transition appears in
Section~\protect\ref{magd}, while that of the transition
describing the onset of $d_{x^2-y^2}+id_{xy}$ order appears in
Section~\protect\ref{qft}. The second transition may not be on the
physical axis for a given cuprate superconductor, but may be
nearby in a generalized parameter space.} \label{fig13}
\end{figure}

We close this subsection by presenting a unified overview of the
quantum phase transitions in the cuprate superconductors we have
introduced so far: this is presented in Fig~\ref{fig13}.
At low, but finite, doping we have the SC+SDW phase which was
discussed in Section~\ref{magd}. Here the presence of the SDW
order induces spatial anisotropy between the $x$ and $y$ axis (in
real space), and so the superconductivity is not purely $d$-wave,
although gapless nodal fermions may be
present~\cite{kwon1,granath}. With increasing $\delta$ we have a
transition to a SC phase, where the superconducting order is
$d_{x^2-y^2}$, and is described by a vanilla BCS theory: this
transition was also discussed in Section~\ref{magd}. Finally, the
results of~\cite{maxim1,maxim2} show that the $d_{xy}$
superconductivity at very low electron density: we have used a
phenomenological model of this crossover in the present section to
argue that this implies that the next phase with increasing doping
should be a $d_{x^2-y^2}+i d_{xy}$ superconductor. The full
critical theory of this transition appears in the following
subsection. In placing the $d_{x^2-y^2}+i d_{xy}$ at large doping,
we are also following the proposal and experimental results of
Dagan and Deutscher~\cite{dagan} which we will discuss further in
Section~\ref{expts}.

\subsection{Quantum field theory}
\label{qft}

We will now present a full theory for the quantum transition
between the $d_{x^2-y^2}$ and $d_{x^2-y^2} + i d_{xy}$
superconductors: this will allow us to compute the corrections to
the predictions of the BCS theory implied by (\ref{g13}).

We have laid much of the ground work for the required field theory
in Section~\ref{magd}. In addition to the order parameter $\phi$,
the field theory should also involve the low energy nodal fermions
of the $d_{x^2-y^2}$ superconductor, as described by
$\mathcal{S}_{\Psi}$ in (\ref{g6}). For the $\phi$ fluctuations,
we write down the usual terms permitted near a phase transition
with Ising symmetry, and similar to those in (\ref{zl1}):
\begin{equation}
\mathcal{S}_{\phi} = \int d^2 r d\tau \left[\frac{1}{2} \left(
(\partial_{\tau} \phi)^2 + c^2 (\partial_x \phi)^2 + c^2
(\partial_y \phi)^2 + s \phi^2 \right) + \frac{u}{24} \phi^4
\right]. \label{g15}
\end{equation}
Note that, unlike (\ref{g13}), we do not have any non-analytic
$|\phi|^3$ terms in the action: this is because we have not
integrated out the low energy nodal fermions, and the terms in
(\ref{g15}) are viewed as arising from high energy fermions away
from the nodal points. Finally, we need to couple the $\phi$ and
$\Psi_{1,2}$ excitations. Their coupling is already contained in
the last term in (\ref{g9}): expressing this in terms of the
$\Psi_{1,2}$ fermions using (\ref{g5a}) we obtain
\begin{equation}
\mathcal{S}_{\Psi \phi} = \int d^2 r d \tau \left[ \lambda \phi
\left( \Psi_1^{\dagger} \tau^y \Psi_1 - \Psi_2^{\dagger} \tau^y
\Psi_2 \right) \right], \label{g16}
\end{equation}
where $\lambda$ is a coupling constant. The partition function of
the full theory is now
\begin{equation}
Z_{did} = \int \mathcal{D}\phi \mathcal{D} \Psi_1 \mathcal{D}
\Psi_2 \exp \left( - \mathcal{S}_{\Psi} - \mathcal{S}_{\phi} -
\mathcal{S}_{\Psi \phi} \right), \label{g17}
\end{equation}
where $\mathcal{S}_{\Psi}$ was in (\ref{g6}). It can now be
checked that if we integrate out the $\Psi_{1,2}$ fermions for a
spacetime independent $\phi$, we do indeed obtain a $|\phi|^3$
term in the effective potential for $\phi$.

We begin our analysis of $Z_{did}$ in (\ref{g17}) by following the
procedure discussed below (\ref{g7}). Assume that the transition
is described by a fixed point with $\lambda=0$: then as in
Section~\ref{magd}, the theory for the transition would be the
ordinary $\phi^4$ field theory $\mathcal{S}_{\phi}$, and the nodal
fermions would again be innocent spectators (we already know from
the presence of the $|\phi|^3$ term in (\ref{g13}) that this
assumption cannot be correct, but we wish to see how it is
invalidated in a renormalization group analysis). The scaling
dimension of $\phi$ at such a fixed point is $(1 + \eta_I )/2$
(where $\eta_I$ is the anomalous order parameter exponent at the
critical point of the ordinary three dimensional Ising model),
while that of $\Psi_{1,2}$ is 1 (as before). Consequently scaling
dimension of $\lambda$ is $(1-\eta_I)/2 > 0$ (using the known,
very small value of $\eta_I$). This positive scaling dimension
implies that $\lambda$ is relevant and the $\lambda=0$ fixed point
is unstable.

Determining the correct critical behavior now requires a full
renormalization group analysis of $Z_{did}$. This has been
described in some detail in~\cite{didrg}, and we will not
reproduce the details here. The main result we need for our
purposes is that couplings $\lambda$, $u$, $v_F /c$ and
$v_{\Delta}/c$ all reach {\em non-zero} fixed point values which
define a critical point in a new universality class. These fixed
point values, and the corresponding critical exponents, can be
determined in expansions in either $(3-d)$~\cite{vojta1,vojta2}
(where $d$ is the spatial dimensionality) or $1/N$~\cite{kvesch}
(where $N$ is the number of fermion species). Indeed the fixed
point has the structure of the so-called Higgs-Yukawa model which
has been studied extensively in the particle physics
literature~\cite{baruch} in a different physical context: quantum
Monte Carlo simulation of this model also exist~\cite{qmc}, and
provide probably the most accurate estimate of the exponents.

The implications of the existence of this finite-coupling critical
point are similar to those for (\ref{zl1}) in (\ref{e4}). The
fermion correlation function $G_1 = \langle \Psi_1
\Psi_1^{\dagger} \rangle$ obeys
\begin{equation}
G_1 (k, \omega) = \frac{\omega + v_F k_x \tau^z + v_{\Delta}
\tau^x}{(v_F^2 k_x^2 + v_{\Delta}^2 k_y^2 -
\omega^2)^{(1-\eta_f)/2}} \label{g18}
\end{equation}
at low frequencies for $s \geq s_c$. Away from the critical point
in the $d_{x^2-y^2}$ superconductor with $s>s_c$, (\ref{g18})
holds with $\eta_f = 0$, and this is the BCS result, with sharp
quasi-particle poles in the Green's function. At the critical
point $s=s_c$ (\ref{g18}) holds with the fixed point values for
the velocities (which satisfy $v_F = v_{\Delta} = c$) and with the
anomalous dimension $\eta_f \neq 0$
--- the $(3-d)$ expansion~\cite{vojta1} estimate is $\eta_f
\approx (3-d)/14$, and the $1/N$ expansion estimate~\cite{kvesch}
is $\eta_f \approx 1/(3 \pi^2 N)$, with $N=2$. This is clearly
non-BCS behavior, and the fermionic quasiparticle pole in the
spectral function has been replaced by a branch-cut representing
the continuum of critical excitations. The corrections to BCS
extend also to correlations of the Ising order $\phi$: its
expectation value vanishes as (\ref{g14}) with the Monte Carlo
estimate $\beta \approx 0.877$~\cite{qmc}. The critical point
correlators of $\phi$ also obey (\ref{e4}) with the exponent $\eta
\approx 0.754$~\cite{qmc}, which is clearly different from the
very small value of the exponent $\eta_I$ at the unstable
$\lambda=0$ fixed point. The value of $\beta$ is related to $\eta$
by the usual scaling law $\beta = (1+\eta) \nu/2$, with $\nu
\approx 1.00$ the correlation length exponent (which also differs
from the exponent $\nu_I$ of the Ising model).

\subsection{Connection with experiments}
\label{expts}

We noted at the beginning of this section that our original
motivation for studying the $d_{x^2-y^2}$ to $d_{x^2-y^2} + i
d_{xy}$ transition was provided by photoemission experiments of
Valla {\em et al.}~\cite{valla}. In particular these experiments
found that the typical quasiparticle energy width was of order
$k_B T$. As we have reviewed elsewhere~\cite{book}, this is a
characteristic property of the $T>0$ quantum critical region of a
$T=0$ critical point which obeys strong hyperscaling properties.
In particular, it is essential that the critical point be
described by an interacting field theory, with the interaction
strengths determined by a finite-coupling fixed point of the
renormalization group. This condition is clearly satisfied by
$Z_{did}$: the breakdown of BCS theory at $s=s_c$, which leads to
(\ref{g18}) with $\eta_f \neq 0$ at $T=0$, is directly responsible
for quasiparticle lifetimes of order $\hbar/k_B T$ over a finite
range of $g$ values in the $T>0$ quantum critical region. In
Ref.~\cite{vojta1} it was also shown that the fixed point of
$Z_{did}$ was also essentially unique in satisfying such
requirements: the only other zero momentum order parameter which
led to suitable critical point was that between a $d_{x^2-y^2}$
and a $d_{x^2-y^2} + is$ superconductor.

A more direct probe of the $d_{x^2-y^2}+i d_{xy}$ order has been
provided by very interesting tunneling measurements of Dagan and
Deutscher~\cite{dagan}. They examined the splitting of a zero-bias
conductance peak as a function of doping $\delta$ and an applied
magnetic field, $H$, perpendicular to the two-dimensional electron
gas. They found that this splitting closely tracked the value
$\langle \phi \rangle$ as determined by minimizing the energy in
(\ref{g13}) extended with a linear coupling to the magnetic field
\begin{equation}
E_{BCS} = E_0 + s \phi^2 + v |\phi|^3 -H \phi. \label{g19}
\end{equation}
The linear coupling is also just that expected from the fact that
the $d_{x^2-y^2}+id_{xy}$ superconductor has a spontaneous orbital
magnetic moment~\cite{orbital,wolf}. The experiments of
Ref.~\cite{dagan} are able to scan across the critical point at
$s=0$, and observe behavior consistent with (\ref{g19}); we regard
this as strong evidence for the phase diagram of Fig~\ref{fig13}
with a quantum critical point between  $d_{x^2-y^2}$ and
$d_{x^2-y^2}+id_{xy}$ superconductors. It would be interesting if
future experiments are able to test for the corrections to
(\ref{g19}) discussed in Section~\ref{qft}. We also note here the
recent measurements of Yeh {\em et al.}~\cite{ncy} presenting
evidence for related (but somewhat different) changes in the
pairing symmetry in the overdoped regime.

\section{Conclusions}

We have studied a number of phase transitions of two dimensional
correlated electron systems and a schematic phase diagram
summarizing their relevance to the high temperature
superconductors was in Fig~\ref{fig13}.  At zero doping,
$\delta=0$, we start from an insulating antiferromagnet with long
range N\'{e}el order. After a localized carrier regime at very low
doping, the ground state at small $\delta$ appears to be
superconducting with co-existing collinear spin-density wave order
(the SC+SDW phase). There is a long in-plane spin correlation
length in this phase (the finiteness of the correlation length is,
we believe, due to disorder effects, and so at long enough scales
the spin order is spin-glass like). At special commensurate
dopings (like $\delta=1/8$) the inter-plane coupling is strong
enough to induce three-dimensional long-range order at finite
temperatures.

At larger $\delta$ there is a transition to a pure $d_{x^2-y^2}$
superconductor, and this transition was described in
Section~\ref{magd}. Charge order and its fluctuations also play a
central role in the magnetic ordering quantum transition: we
discussed these effects in the insulator in Section~\ref{2d}. They
also have an important effect near the corresponding SC+SDW to SC
transition in the superconductor, but the reader is referred
to~\cite{sns} for a review of these.

With increasing $\delta$, we argued on the basis of
theoretical~\cite{maxim1,maxim2} and experimental
studies~\cite{dagan} that the cuprate superconductors are at least
in the vicinity of a separate transition from the $d_{x^2-y^2}$
superconductor to a $d_{x^2-y^2}+id_{xy}$ superconductor. A
quantum field theory for this transition, associated with the
breakdown of BCS theory at the critical point, was presented in
Section~\ref{did}.

\section*{Acknowledgements} I am grateful to the organizers, and
especially Bernard Nienhuis, for all their hard work, and the
participants for their interest. This research was supported by US
NSF Grant DMR 0098226.

\end{document}